\documentclass[conference]{IEEEtran}
\usepackage{epsfig}
\usepackage{graphicx}
\usepackage{amsmath}
\usepackage{amssymb}
\usepackage{amsxtra}
\usepackage{here}
\usepackage{rawfonts}
\usepackage{times}
\usepackage{url}
\usepackage{cite}

\usepackage{algorithm}

\usepackage{algorithmic}

\usepackage{ dsfont }

\usepackage{cases}
\usepackage{adjustbox}

\usepackage[usenames, dvipsnames]{color}

\usepackage{slashbox}
\usepackage{epstopdf}

\usepackage{cite}
\newtheorem{prop}{\bf Proposition}

\newtheorem{lemma}{\bf Lemma}
\newtheorem{definition}{\bf Definition}
\newtheorem{corollary}{\bf Corollary}
\newtheorem{theorem}{\bf Theorem}
\newtheorem{example}{\bf Example}

\newlength{\aligntop}
\newlength{\alignbot}
\makeatletter

\renewenvironment{align}{%
  \vspace{\aligntop}
  \start@align\@ne\st@rredfalse\m@ne
}{%
  \math@cr \black@\totwidth@
  \egroup
  \ifingather@
    \restorealignstate@
    \egroup
    \nonumber
    \ifnum0=`{\fi\iffalse}\fi
  \else
    $$%
  \fi
  \ignorespacesafterend%
  \vspace{\alignbot}\par\noindent
}

\IEEEoverridecommandlockouts

\begin{document}
\title{\LARGE Graph-Theoretic Framework for Unified Analysis of Observability and Data Injection Attacks in the Smart Grid\vspace{-0.4cm}}
\author{\IEEEauthorblockN{Anibal Sanjab$^{1,2}$, Walid Saad$^1$, and Tamer Ba\c{s}ar$^3$} \IEEEauthorblockA{\small
$^1$ Wireless@VT, Bradley Department of Electrical and Computer Engineering, Virginia Tech, Blacksburg, VA, USA,\\
 Emails: \url{{anibals,walids}@vt.edu}\\
$^2$ Flemish Institute for Technological Research, VITO/EnergyVille, Genk, Belgium, Email: \url{anibal.sanjab@vito.be}\\
 $^3$ Coordinated Science Laboratory, University of Illinois at Urbana-Champaign, IL, USA, Email: \url{basar1@illinois.edu}\vspace{-0.6cm}
 }%
\thanks{This research was supported by the U.S. National Science Foundation under Grants ECCS-1549894 and CNS-1446621, and in part by the Office of Naval Research (ONR) MURI Grant N00014-16-1-2710.}
    }
\date{}
\maketitle

\begin{abstract}
In this paper, a novel graph-theoretic framework is proposed to generalize the analysis of a broad set of security attacks, including observability and data injection attacks, that target the state estimator of a smart grid. First, the notion of observability attacks is defined based on a proposed graph-theoretic construct. In this respect, a structured approach is proposed to characterize critical sets, whose removal renders the system unobservable.  
It is then shown that, for the system to be observable, these critical sets must be part of a maximum matching over a proposed bipartite graph. In addition, it is shown that stealthy data injection attacks (SDIAs) constitute a special case of these observability attacks. Then, various attack strategies and defense policies, for observability and data injection attacks, are shown to be amenable to analysis using the introduced graph-theoretic framework. 
The proposed framework is then shown to provide a unified basis for analysis of four key security problems (among others), pertaining to the characterization of: 1) The sparsest SDIA; 2) the sparsest SDIA including a certain measurement; 3) a set of measurements which must be defended to thwart all potential SDIAs; and 4) the set of measurements, which when protected, can thwart any SDIA whose cardinality is below a certain threshold.   
A case study using the IEEE 14-bus system with a set of $17$ measurements is used to support the theoretical findings.                    
\end{abstract}
\section{Introduction}\label{sec:intro} 

With the integration of information and communication technologies in power systems, new security concerns have emerged due to the potential exploitation of this cyber layer to infiltrate and compromise the underlying physical system. Indeed, in recent years, various studies have focused on analyzing the security of emerging cyber-physical power systems~\cite{SGSecSurvey1,liu1stdatainjectionConf,SanjabSaadJournal,SanjabSaad,LeXie,AttDetDorfler,RobustResilientCPSPwrSysBasar} and the effect of potential cyber attacks on the various operational components of the grid, ranging from power system state estimation~\cite{liu1stdatainjectionConf}, to electricity markets~\cite{SanjabSaadJournal,SanjabSaad,LeXie} and power system dynamics and control~\cite{AttDetDorfler,RobustResilientCPSPwrSysBasar}. 

Such attacks can become more pronounced when they target critical power system functions such as state estimation. In this regard, the power system state estimation is an integral smart grid process in which system-wide measurements are collected and processed to estimate the global state of operation of a power system~\cite{Abur}. State estimation is the basis for various grid operational decisions such as congestion management, economic dispatch, contingency analysis, and electricity pricing~\cite{woodwollenberg}. As a result, the critical importance of state estimation to the sustainable operation of the grid makes it a primary target of possible cyber-physical attacks~\cite{SGSecSurvey1}. Such attacks may target the availability of the collected measurements as well as their integrity. 

In this respect, intercepting a subset of the collected measurement data using availability attacks (such as denial-of-service attacks) can render the power system unobservable (i.e. not fully observable), a state in which the collected measurements do not provide enough independent equations to estimate the states. Such cyber-physical attacks, to which we refer as \emph{observability attacks} hereinafter, will make the operator partially oblivious to the real state of operation of the system, leading to uninformed operational decisions. Beyond observability attacks, \emph{data injection attacks} (DIAs) have emerged as a malicious type of integrity attacks which aim at manipulating the collected state estimation data, leading to inaccurate state estimation outcomes that result in misinformed operational decisions with potentially detrimental consequences~\cite{SGSecSurvey1,liu1stdatainjectionConf,SanjabSaadJournal}. As shown in~\cite{liu1stdatainjectionConf}, such DIAs can stealthily target the power system state estimation process -- manipulating the collected measurements and altering the state estimation outcome -- while being undetectable by the system operator using traditional bad data detection mechanisms. Hence, due to their potential danger to system operation, such stealthy data injection attacks (SDIAs) and observability attacks have been the focus of various recent research efforts~\cite{CritSet3,DataInjGTKTuple,DataInjGTKosut,DataInjSecIndex,SparsestAttackCDC,DIAsSecIndexTAC,Poor}. 

\subsection{Related Works}
In this regard, the works in~\cite{CritSet3} and~\cite{DataInjGTKTuple} focused on computing a security set which comprises the minimum set of measurements which must be attacked in addition to a certain specific measurement in order to make the system unobservable. Moreover, the work in~\cite{DataInjGTKosut} focused on computing the cardinality of the smallest set of meters which when attacked render the system unobservable. The authors in~\cite{DataInjSecIndex,SparsestAttackCDC,DIAsSecIndexTAC} extended such observability problems to studying SDIAs. In this regard, these works focused on characterizing the sparsest stealthy attack containing a certain specific measurement. In addition, the work in~\cite{Poor} focused on characterizing a set of measurements to defend so that no attack which concurrently manipulates a set of meters whose cardinality is below a certain threshold can be stealthy. 
Hence, this latter analysis focuses on the defense against resource-limited attackers. As such, these works have focused on formulating and studying mathematical problems whose solutions enable anticipating potential sophisticated attacks -- which constitutes a first step towards deriving corresponding defense mechanisms -- and designing optimal defense strategies to thwart such attacks and mitigate their potential effect. 

The computational complexity of these problems~\cite{CritSet3,DataInjGTKTuple,DataInjGTKosut,DataInjSecIndex,SparsestAttackCDC,DIAsSecIndexTAC,Poor} has led to limiting the analysis of their solutions to special, often approximated, cases or required the use of heuristics and relaxation techniques which led to suboptimal solutions. For example, for characterizing the sparsest observability attacks containing a specific measurement, the work in~\cite{CritSet3} focused on the special case of measurement sets of low cardinality while the work in~\cite{DataInjGTKTuple} derived an approximate solution that is based on the solution of a min-cut problem. In addition, with regard to the analysis of the sparsest SDIAs containing a certain measurement~\cite{DataInjSecIndex,SparsestAttackCDC,DIAsSecIndexTAC}, the work in~\cite{DataInjSecIndex} focused on deriving an upper-bound on this stealthy attack set while the work in~\cite{SparsestAttackCDC} used min-cut relaxation techniques to approximate the sought solution. Moreover, the work in~\cite{DIAsSecIndexTAC} proposed a heuristic algorithm which can approximate the solution of the studied problem while an exact solution was found for the special case in which power flows over all the transmission lines and power injections into and out of every bus are assumed to be measured. To defend against a resource-limited data injection attacker, the authors in~\cite{Poor} used an $l_1$ relaxation method for characterizing the set of meters to defend to thwart SDIAs launched by attackers whose attack space is limited by a certain cardinality threshold. Other related security works are also found in~\cite{TajerPoor,CritSet3Journal,MminusKRobustEstimation,SparsestSDIAOnlyFlowMeas,MinSparcityDIAPMUTAC,MinSparcityDIAPMUACC}.   

Therefore, this rich body of literature~\cite{CritSet3,DataInjGTKTuple,DataInjGTKosut,DataInjSecIndex,SparsestAttackCDC,DIAsSecIndexTAC,Poor,TajerPoor,CritSet3Journal,MminusKRobustEstimation,SparsestSDIAOnlyFlowMeas,MinSparcityDIAPMUTAC,MinSparcityDIAPMUACC} employs heuristics and approximation techniques to numerically approximate the solutions to these fundamental observability attacks and SDIA problems rather than propose new analytical methods for studying these problems and characterizing their solutions. As such, there is a need for an analytical framework which allows modeling and studying such data availability and integrity attacks and enables an analytical characterization of solutions to such widely-studied security problems. In addition, the fact that these works~\cite{CritSet3,DataInjGTKTuple,DataInjGTKosut,DataInjSecIndex,SparsestAttackCDC,DIAsSecIndexTAC,Poor} studied correlated problems but from different perspectives highlights the need for a unified framework using which solutions to such correlated observability attacks and SDIA problems can be studied and derived. 
  
\subsection{Contributions}
The main contribution of this paper is a novel unified graph-theoretic framework that enables a global detailed modeling and understanding of observability attacks and SDIAs. As a result, this framework provides a unified tool for analyzing various widely-studied observability attacks and SDIA problems such as those studied in~\cite{CritSet3,DataInjGTKTuple,DataInjGTKosut,DataInjSecIndex,SparsestAttackCDC,DIAsSecIndexTAC,Poor}, among others. In addition, the proposed framework enables a graph-theoretic characterization of solutions to such security problems. 
In this regard, our proposed framework is based on a shift in the modeling of observability attacks and SDIAs from a linear algebra frame of reference to a graph-theoretic perspective. As a result, based on this proposed framework, such attacks can be modeled and analyzed by requiring only power system topological data, namely, the power system 1-line diagram and the location of deployed measurement units without the need for neither line parameters data nor the exact knowledge of power flow levels throughout the system. 
%
%

To build the proposed framework, we first begin by introducing a graph-theoretic basis of observability attacks and, then, we prove that SDIAs are a special case of such observability attacks. In this respect, we introduce an algorithm providing a step-by-step approach for building \emph{critical sets}, a set of measurements -- containing a certain specific measurement -- which, when removed, renders the system unobservable. We then prove that for a DIA to be stealthy, the attacked measurements should strictly result in leaving critical sets unmatched as part of a maximum matching over an introduced bipartite graph. As such, a graph-theoretic model of SDIAs is then introduced based on which we analyze various well-studied SDIA problems. In particular, \emph{we show that our developed framework enables a graph-theoretic characterization of solutions to various SDIA problems} such as, but not limited to: 1) Finding the stealthy attack of lowest cardinality, 2) Finding the stealthy attack of lowest cardinality, including a specific measurement, 3) Finding a set of measurements which when defended can thwart all possible stealthy attacks, and 4) Finding a set of measurements to defend against a resource-limited attacker, among others. Here, we note that our goal is not to propose tractable algorithms to these problems, but rather to identify graph-theoretic problems whose solutions would lead to the solutions of these problems. 
A case study using the IEEE 14-bus system, with $17$ distributed measurement units, is considered throughout the paper to showcase the developed analytical concepts. 

The rest of the paper is organized as follows. Section~\ref{sec:SE} introduces state estimation and power system observability. Section~\ref{sec:ObservabilityAtt} introduces our proposed graph-theoretic foundation of observability attacks and shows its impact on modeling and analyzing such data availability attacks. Section~\ref{sec:StealthyDIAs} introduces the proposed graph-theoretic framework for modeling SDIAs, as well as investigates various well-studied SDIA problems. Section~\ref{sec:Conclusion} concludes the paper and provides an outlook detailing the impact of the proposed framework on studying future observability and data injection attacks. 

A summary of the main notations used in this paper is given in Table~\ref{Tab:Notation}.
\begin{table}[t!]
	\caption{Summary of main notations.}\label{Tab:Notation}\vspace{-0.4cm}
	\begin{center}
		\begin{tabular}{|c|c|}
		\hline
                   $\boldsymbol{z}\in\mathds{R}^{\mu}$& measurement vector with $\mu$ measurements \\ \hline
                   $\boldsymbol{x}\in\mathds{R}^{\nu}$& state vector with $\nu$ states \\ \hline
                   $\boldsymbol{H}\in\mathds{R}^{\mu\times \nu}$ & system's Jacobian matrix\\ \hline
                    $\mathcal{G}(\mathcal{N},\mathcal{L})$ & power system graph with $N$ buses \& $L$ lines  \\ \hline  
                    $\mathcal{M}$ & set of measurements in $\mathcal{G}(\mathcal{N},\mathcal{L})$ \\ \hline 
                    $\mathcal{T}(\mathcal{N},\mathcal{B})$ & spanning tree over $\mathcal{G}(\mathcal{N},\mathcal{L})$ \\ \hline                  
                    $\mathcal{M}^A\subseteq\mathcal{M}$& set of assigned measurements  \\ \hline    
                    $f(.)$$:$ $\mathcal{M}^A\rightarrow \mathcal{B}$ & measurement assignment function\\ \hline 
                    $\mathcal{C}^m$ & critical set of measurement $m$ \\ \hline
                    $\mathcal{T}^m_i(\mathcal{N}^m_i,\mathcal{B}^m_i)$ & spanning tree over subgraph $\mathcal{G}^m_i(\mathcal{N}^m_i,\mathcal{L}^m_i)$\\ \hline    
                    $\mathcal{M}^m_i$ & set of measurements in subgraph $\mathcal{G}^m_i(\mathcal{N}^m_i,\mathcal{L}^m_i)$\\ \hline 
                    $\mathcal{L}^m$ & set of lines connecting subgraphs $\mathcal{G}^m_1$ and $\mathcal{G}^m_2$\\ \hline 
                    $\mathcal{M}^m$ & \!set of measurements on or incident to $\mathcal{L}^m$ excluding $m$\\ \hline  
		\end{tabular}
	\end{center}\vspace{-0.6cm}
\end{table}

\section{State Estimation and Observability}\label{sec:SE}
We next provide an overview of state estimation and of the algebraic and topological concepts of observability in power systems. This overview provides background material which is useful for the analysis that follows. 

\subsection{State Estimation Process}\label{subsec:SEProcess}
Consider a power system state estimation process which uses various measurements collected from across the system to estimate the voltage magnitudes and phase angles at every bus in the system, known as the system states~\cite{Abur}. Let $\boldsymbol{z}\in\mathds{R}^\mu$ ($\mu$ being the number of measurements) be the vector of collected measurements, which includes power flow levels (real and reactive) over transmission lines, power (real and reactive) injected in or withdrawn from certain buses, as well as bus voltage magnitudes. 
In addition, let $\boldsymbol{x}\in\mathds{R}^\nu$ be the vector of system states. The relationship between the measurements and the states directly follows from the linearized power flow equations~\cite{Abur}:
\begin{align}\label{eq:MeasState}
\boldsymbol{z}=\boldsymbol{H}\boldsymbol{x}+\boldsymbol{e}, 
\end{align}     
where $\boldsymbol{H}\in\mathds{R}^{\mu\times \nu}$ is the measurement Jacobian matrix and $\boldsymbol{e}\in\mathds{R}^{\mu}$ is the vector of random errors that typically follows a Gaussian distribution, ${N(0,\boldsymbol{R})}$, where $\boldsymbol{R}$ is positive definite. Here $\mu\geq \nu$, that is the dimension of $\boldsymbol{x}$ cannot be larger than the dimension of the measurement vector, $\boldsymbol{z}$. Further, we assume that $\boldsymbol{H}$ is a full column rank matrix. 
%
Using a maximum-likelihood estimator -- a weighted least squares estimator (WLS) for a Gaussian error vector $\boldsymbol{e}$ -- an estimate of the states, $\hat{\boldsymbol{x}}$, will be:
\begin{align}\label{eq:SE}
\boldsymbol{\hat{x}}=(\boldsymbol{H}^T\boldsymbol{R}^{-1}\boldsymbol{H})^{-1}\boldsymbol{H}^T\boldsymbol{R}^{-1}\boldsymbol{z}. 
\end{align}  

This estimate of all the states provides visibility of the steady-state operating conditions of the system, based on which various operational decisions are performed~\cite{Abur}.

\subsection{Power System Observability}\label{subsec:Observability} 

The observability\footnote{Our analysis focuses on static state estimation and static observability of the power system, which is a fundamental aspect of power system analysis~\cite{Abur}. Our analysis, thus, does not extend to dynamic observability of a power system, which would arise with dynamically changing states.} of the power system consists of the ability to uniquely determine its states based on the collected set of measurements~\cite{Abur}. Observability, hence, requires the collected measurements to provide a sufficient number of independent equations to allow for the estimation of the state vector, $\boldsymbol{x}$. Otherwise, when this observability condition is not met, the power system is dubbed \emph{unobservable}. The power system is \emph{observable} if and only if the measurement matrix $\boldsymbol{H}$ is of full column rank~\cite{Abur}, which was our initial assumption. 
This is known as \emph{algebraic observability}, as it enables assessment of observability using linear algebra. Due to the $P-\theta$, $Q-V$ decoupling\footnote{$P$ denotes real power, $\theta$ denotes voltage phase angles, $Q$ denotes reactive power, and $V$ denotes voltage magnitudes.} in power systems~\cite{woodwollenberg}, the observability analysis can be decoupled by separately studying the observability of voltage phase angles, using real power measurements, and the observability of voltage magnitudes, based on reactive power measurements. Since the two analyses are identical, we focus here on phase angle observability. To this end, we consider $\boldsymbol{z}\in\mathds{R}^{\mu}$ to be a vector of real power measurements (bus injections and line flows), and the state vector $\boldsymbol{x}\in [-\pi,\pi]^{\nu}$ to be the vector of voltage phase angles (in radians). Here, $\nu=N-1$ for a power system with $N$ buses given that the phase angle of the reference bus is fixed and is taken to be the reference with respect to which all other phase angles are calculated~\cite{Abur}.   

An alternative measure of observability, which turns out to be equivalent to algebraic observability, has been proposed in~\cite{TopologicalObs1} and uses graph-theoretic techniques to introduce the concept of \emph{topological observability}. Topological observability is equivalent to algebraic observability, in the sense that it enables assessment of the observability of the power system using graph-theoretic tools rather than using linear algebra as done for algebraic observability. In this regard, let the power system 1-line diagram be represented as a graph $\mathcal{G}(\mathcal{N},\mathcal{L})$ in which the set of vertices $\mathcal{N}$, $|\mathcal{N}|=N$, represents the set of buses of the power system while the set of branches $\mathcal{L}$, $|\mathcal{L}|=L$, represents the set of lines. 
One key result that was shown in~\cite{TopologicalObs1} and that will be of relevance to our work is the following:
\begin{prop}[\!\!\cite{TopologicalObs1}\,]\label{remark:TopologicalObs}
A power system is \emph{observable} if and only if a subset of measurements can be assigned to a subset of edges of the power system graph $\mathcal{G}(\mathcal{N},\mathcal{L})$, following a set of assignment rules, in a way to form a \emph{spanning tree} over $\mathcal{G}(\mathcal{N},\mathcal{L})$. A spanning tree over $\mathcal{G}(\mathcal{N},\mathcal{L})$ is an acyclic connected subgraph of $\mathcal{G}(\mathcal{N},\mathcal{L})$ which contains (i.e. is incident to) the entire set of nodes $\mathcal{N}$ of $\mathcal{G}(\mathcal{N},\mathcal{L})$.
The set of measurement assignment rules are the following:
\begin{enumerate}
\item{A measurement cannot be simultaneously assigned to two different lines.} 
\item{If $m$ is a measurement over a transmission line $l$, then $m$ can only be assigned to $l$.} 
\item{If $m$ is an injection measurement over bus $\eta\in\mathcal{N}$, then $m$ can only be assigned to an unmeasured line $l$ that is incident to $\eta$.} 
\end{enumerate}
\end{prop}



\begin{figure}[t!]
  \begin{center}
   \vspace{-0.35cm}
    \includegraphics[width=8cm]{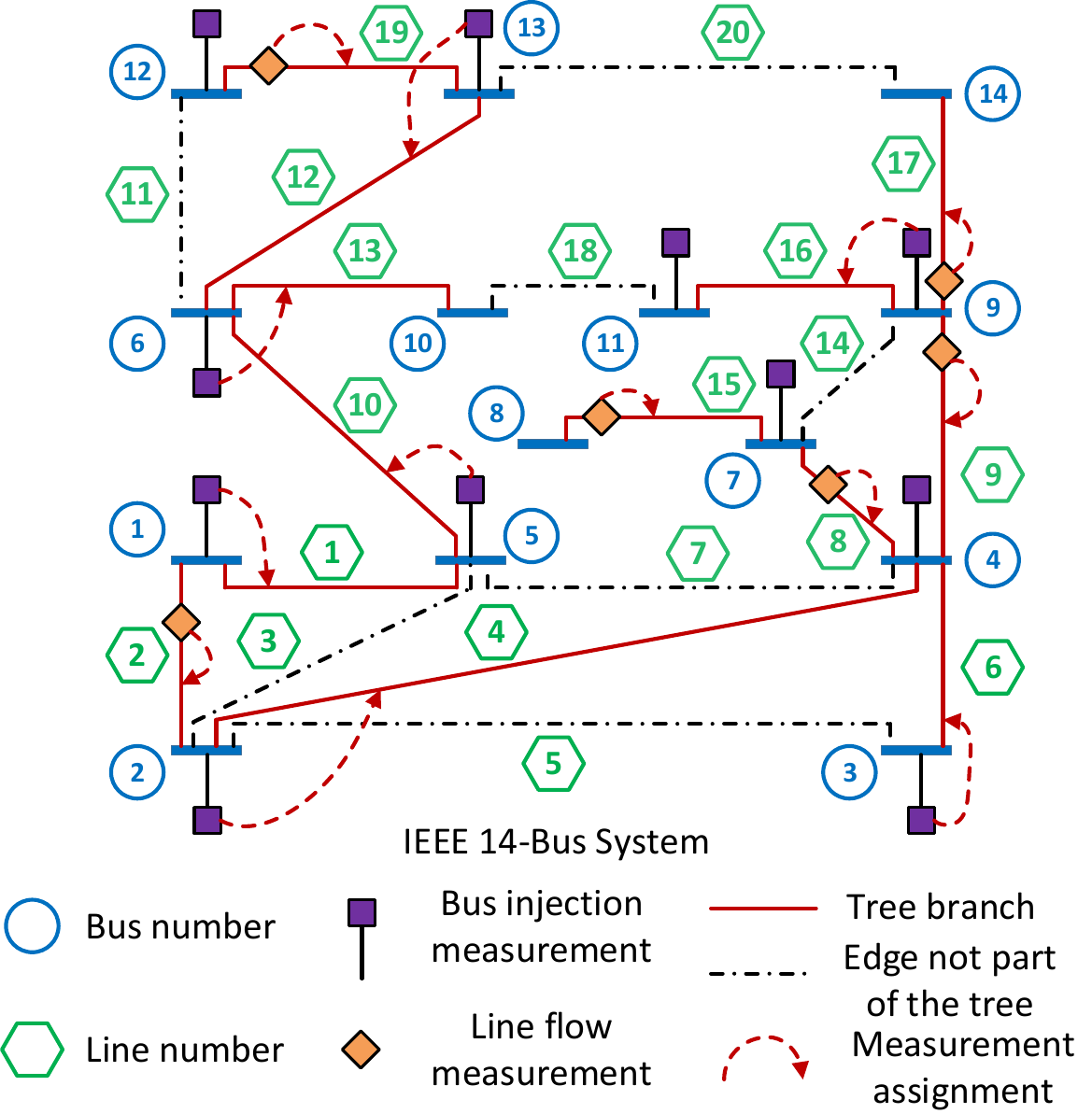}
    \vspace{-0.4cm}
    \caption{\label{fig:IEEE14Bus} IEEE 14-bus system with measurement assignment.}
  \end{center}\vspace{-0.6cm}
\end{figure}

If a measurement assignment yields a spanning tree over the power network $\mathcal{G}$, then the power system will be observable (and vice versa). 
Fig.~\ref{fig:IEEE14Bus} shows an example of measurement assignments over the IEEE 14-bus system. This figure shows the tree edges (marked in solid red lines) to which measurement where assigned as part of the measurement assignment function. The measurements that were assigned to each one of these edges are identified using dashed arrow lines originating from the assigned measurement and pointing to the line to which this measurement is assigned. This tree is formed of edges $\{1,\,2,\,4,\,6,\,8,\,9,\,10,\,12,\,13,\,15,\,16,\,17,\,19\}$ and spans the whole vertex set $\mathcal{N}$ of the power system graph $\mathcal{G}$, and hence, is a spanning tree. As a result, since this measurement assignment yields a spanning tree, then the available set of measurements renders the system observable.  

Let $\mathcal{M}$ be the set of measurements, and let $\mathcal{T}(\mathcal{N},\mathcal{B})$ be a spanning tree formed by an assignment of a subset of measurements, $\mathcal{M}^A\subseteq \mathcal{M}$, to a subset of lines $\mathcal{B}\subseteq\mathcal{L}$ following the measurement assignment rules in Proposition~\ref{remark:TopologicalObs}. The assignment of measurements in $\mathcal{M}^A$ to lines in $\mathcal{B}$ for constructing the spanning tree $\mathcal{T}(\mathcal{N},\mathcal{B})$ can be modeled by an assignment function $f(.)$ formally defined~\cite{TopologicalObs1} as follows:
\begin{definition}
For an assigned set of measurement $\mathcal{M}^A$ forming a spanning tree $\mathcal{T}(\mathcal{N},\mathcal{B})$ over the power system graph $\mathcal{G}(\mathcal{N},\mathcal{L})$, the assignment function $f(.):\,\mathcal{M}^A\rightarrow \mathcal{B}$ indicates which measurement $m\in\mathcal{M}^A$ is assigned to which edge $b\in\mathcal{B}$ of the spanning tree $\mathcal{T}(\mathcal{N},\mathcal{B})$. 
\end{definition}

For example, in Fig.~\ref{fig:IEEE14Bus}, $\mathcal{M}^A$ is composed of the injection measurements over buses $1,2,3,5,6,9$, and $13$ and the line flow measurements over lines $2, 8,9,15,17$, and $19$. In addition, $\mathcal{B}=\{1,\,2,\,4,\,6,\,8,\,9,\,10,\,12,\,13,\,15,\,16,\,17,\,19\}$, where each edge of $\mathcal{B}$ in Fig.~\ref{fig:IEEE14Bus} is colored in red. Moreover, the corresponding assignment function $f(.)$ is visualized in Fig.~\ref{fig:IEEE14Bus} by dashed red arrows indicating which measurement in $\mathcal{M}^A$ is assigned to which line in $\mathcal{B}$.

Measurements in $\mathcal{M}$ that are not part of $\mathcal{M}^A$, i.e. $m\in\mathcal{M}\setminus{\mathcal{M}^A}$, are not assigned measurements and are, therefore, redundant measurements with respect to the assigned set of measurements $\mathcal{M}^A$. Here, we note that due to this redundancy, a spanning tree which can be obtained from the measurement set following the measurement assignment rules may not be unique~\cite{TopologicalObs1}. Hence, when such a redundancy exists, following a different assignment function can result in a different spanning tree and, hence, in different sets of assigned and redundant measurements.

Various algorithms of low complexity have been proposed to find and build such a spanning tree~\cite{TopologicalObs1,ObservabulityAlgorithmGTMatroid,ObservabilityGTAlgorithm}. 
In this regard, the work in~\cite{TopologicalObs1} proposes an algorithm to find a spanning tree over $\mathcal{G}$, which will be used in some of the derivations in the following sections. This algorithm starts by processing flow measurements by assigning each flow measurement to its corresponding branch to form disjoint tree components. Then, injection measurements are assigned to lines in a way to connect these tree components to form one spanning tree. Here, we highlight one type of injection measurements, namely, \emph{boundary injections}, which will play a crucial role in our derivations. 
\begin{definition}
A boundary injection is an injection measurement over a bus incident to lines whose flow is measured and lines whose flow is not measured~\cite{TopologicalObs1}. 
\end{definition}

Boundary injections play a major role in connecting these tree components. Indeed, for a bus which is not incident to a measured line to be connected to the spanning tree, it has to be reachable from a boundary injection through a series of measurement assignments~\cite{TopologicalObs1}. As such, boundary injections are considered to be sources and unmeasured buses are considered to be sinks which must be connected to these sources following the set of measurement assignment rules.

We next build on the foundation of topological observability to present a graph-theoretic framework for modeling and studying the security of the smart grid facing observability attacks and SDIAs. 
This framework is based on our proposed concepts of critical sets and observability sets, which we define and derive in the next section. 

\section{Observability Attacks}\label{sec:ObservabilityAtt}                                                        

The sustainable and efficient operation of the power system requires an accurate observability of all its states~\cite{Abur}. Security attacks that target this observability can cause a limited (or partial) monitoring ability for the operator over the power system which can lead to incorrect operational decisions. Hence, studying and modeling attacks which can target the full observability of the system is indispensable to the sustainable operation of the grid. In this respect, we define a cyber-physical attack, dubbed \emph{observability attack}, that consists of launching a denial-of-service (DoS) attack against a set of measurements to make the system unobservable. 
We next study this type of attacks by introducing and characterizing what we define as critical sets and observability sets 
and prove that the well-studied stealthy data injection attack is a subset of our defined observability attacks. This latter finding will provide us with a unified set of tools to characterize solutions to various widely-studied SDIA problems.

\subsection{Critical Sets}\label{subsec:CriticalMeas}  
Understanding and modeling observability attacks requires an in-depth understanding of the effect of the loss of any bundle of measurements on the observability of the system. In this regard, we next introduce a structured method for identifying, for each measurement $m$, a minimal set of measurements including $m$ (which we refer to as a critical set of $m$), which when removed renders the system unobservable. To this end, we first characterize the set of potential measurements to be investigated, for each measurement $m$, and then provide the necessary discussion and introduce the underlying method for characterizing a critical set of $m$. 
Then, a detailed algorithm is introduced to provide a step-by-step method for characterizing such critical sets.    

As defined in Section~\ref{subsec:Observability}, we let $\mathcal{M}^A\subseteq\mathcal{M}$ be a set of assigned measurements, i.e., a subset of measurements that are assigned to a subset of lines of the power system graph $\mathcal{G}(\mathcal{N},\mathcal{L})$, following the measurement assignment rules in Proposition~\ref{remark:TopologicalObs}, to form a spanning tree, $\mathcal{T}(\mathcal{N},\mathcal{B})$, over this graph. As described in Section~\ref{subsec:Observability}, $\mathcal{T}(\mathcal{N},\mathcal{B})$ is a spanning tree as it includes all the nodes of $\mathcal{G}(\mathcal{N},\mathcal{L})$. The set of edges of this tree, i.e. $\mathcal{B}$, represents the set of lines to which measurements in $\mathcal{M}^A$ were assigned to construct the spanning tree. We refer to measurements that are not part of $\mathcal{M}^A$ as \emph{unassigned measurements}.  
We consider that the system is originally observable. Hence, such a spanning tree and its corresponding set of assigned measurements exist. 
 
In this regard, consider a spanning tree $\mathcal{T}(\mathcal{N},\mathcal{B})$ resulting from a measurement assignment and consider an assigned measurement $m\in\mathcal{M}^A$. Since $m$ is assigned to a line $f(m)$ to build the spanning tree $\mathcal{T}(\mathcal{N},\mathcal{B})$, its removal will split the original tree $\mathcal{T}(\mathcal{N},\mathcal{B})$ into two trees $\mathcal{T}_1^m(\mathcal{N}^m_1,\mathcal{B}_1^m)$ and $\mathcal{T}_2^m(\mathcal{N}_2^m,\mathcal{B}_2^m)$ each spanning, respectively, subgraphs $\mathcal{G}^m_1(\mathcal{N}^m_1,\mathcal{L}^m_1)$ and $\mathcal{G}^m_2(\mathcal{N}^m_2,\mathcal{L}^m_2)$, such that $\mathcal{N}=\mathcal{N}^m_1\cup\mathcal{N}^m_2$ and $\mathcal{B}=\mathcal{B}^m_1\,\cup\,\mathcal{B}_2^m\,\cup\, \{f(m)\}$. We let $N^m_1=|\mathcal{N}^m_1|$ and $N^m_2=|\mathcal{N}^m_2|$. We refer to the set of measurements within each of  $\mathcal{G}^m_1(\mathcal{N}^m_1,\mathcal{L}^m_1)$ and $\mathcal{G}^m_2(\mathcal{N}^m_2,\mathcal{L}^m_2)$ by, respectively, $\mathcal{M}^m_1$ and $\mathcal{M}^m_2$. In other words, $\mathcal{M}^m_i$, for $i\in\{1,2\}$, consists of injection measurements over buses in $\mathcal{N}^m_i$ and power flow measurements over lines in $\mathcal{L}^m_i$. 

\begin{figure}[t!]
  \begin{center}
   \vspace{-0.35cm}
    \includegraphics[width=9cm]{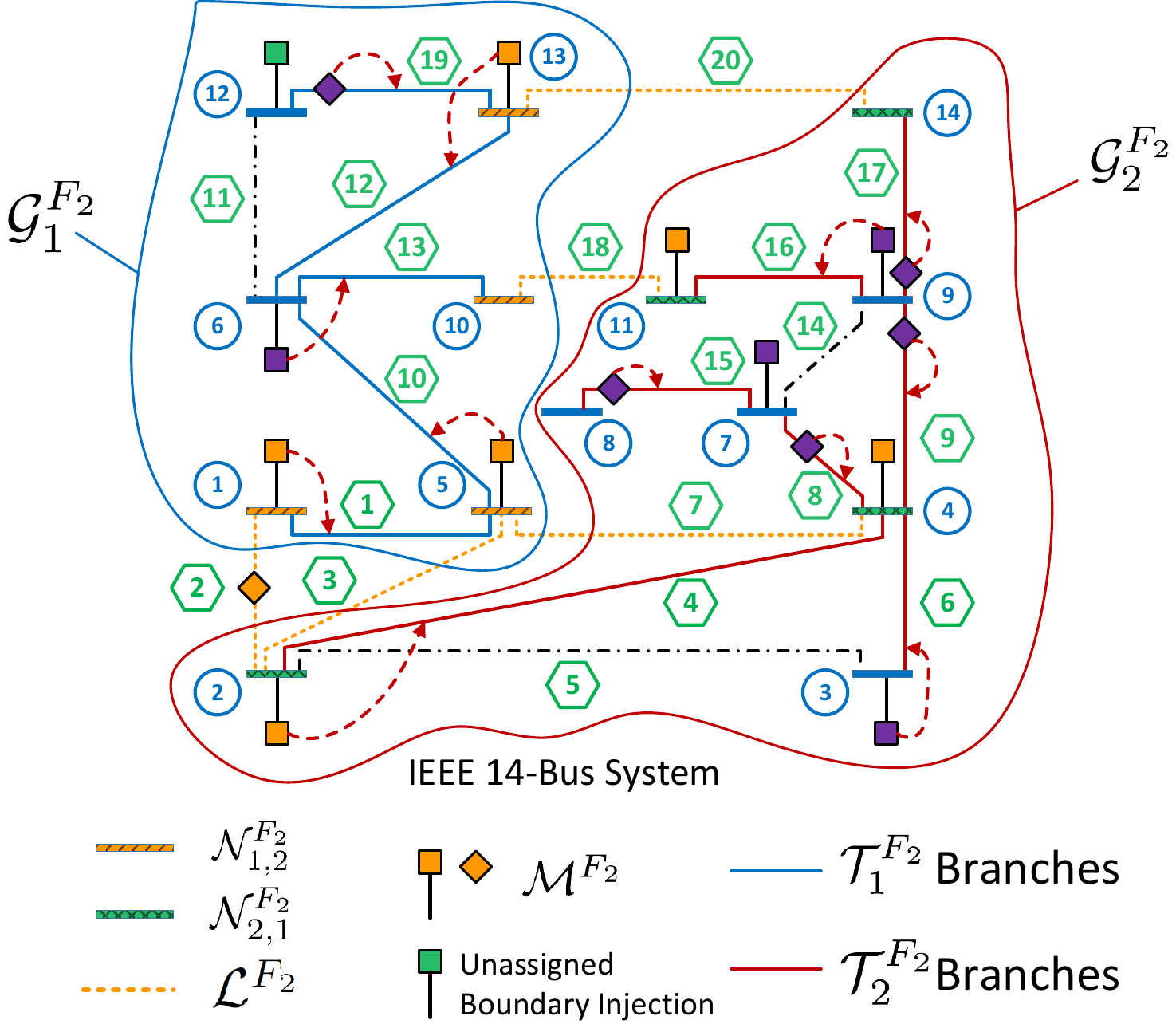}
    \vspace{-0.4cm}
    \caption{\label{fig:IEEE14BusSystemCriticalSetExample} Effect of removal of the flow measurement over line 2, $F_2$, in the IEEE 14-bus system.}
  \end{center}\vspace{-0.65cm}
\end{figure}

Fig.~\ref{fig:IEEE14BusSystemCriticalSetExample} provides an illustrative example of the two spanning trees created by the deletion of flow measurement $F_2$ over line $2$. Here, $f(F_2)$ is line $2$. Fig.~\ref{fig:IEEE14BusSystemCriticalSetExample} represents the same system shown in Fig.~\ref{fig:IEEE14Bus} and will be used hereinafter to provide a practical example of the defined concepts and analytical derivations. 
As can be seen from Fig.~\ref{fig:IEEE14BusSystemCriticalSetExample}, since $F_2$ was assigned to line $2$, when the measurement assignment is not modified (i.e. not considering the redundant measurements), the removal of $F_2$ will split the original tree $\mathcal{T}(\mathcal{N},\mathcal{B})$, represented in solid red lines in Fig.~\ref{fig:IEEE14Bus}, into two trees each of which spans a subgraph of $\mathcal{G}(\mathcal{N},\mathcal{B})$, namely, subgraphs $\mathcal{G}^{F_2}_1$ and $\mathcal{G}^{F_2}_2$ in Fig.~\ref{fig:IEEE14BusSystemCriticalSetExample}.

We let $\mathcal{L}^m$ be the set of lines connecting a bus in $\mathcal{G}_1^m$ to a bus in $\mathcal{G}_2^m$. In addition, let $\mathcal{N}_{1,2}^m\subseteq\mathcal{N}_1^m$ and $\mathcal{N}_{2,1}^m\subseteq\mathcal{N}_2^m$ be the set of nodes in, respectively, $\mathcal{N}_1^m$ and $\mathcal{N}_2^m$ which are connected to a node in, respectively, $\mathcal{N}_2^m$ and $\mathcal{N}_1^m$. An example of these notations is provided in Fig.~\ref{fig:IEEE14BusSystemCriticalSetExample}, for $m\triangleq F_2$. $\mathcal{L}^m$ is, hence, formally defined as:
\begin{align}\label{eq:mLinesConnect}
\mathcal{L}^m=\{l\in\mathcal{L}\,|\,l=(\eta_1,\eta_2), \eta_1\in\mathcal{N}_{1,2}^m, \eta_2\in\mathcal{N}_{2,1}^m\}.
\end{align}  

$\mathcal{G}_1^m$ and $\mathcal{G}_2^m$ are two disjoint subgraphs of $\mathcal{G}$. Hence, for any tree to potentially span $\mathcal{G}$, it must connect $\mathcal{G}_1^m$ and $\mathcal{G}_2^m$ (i.e. connect at least one node of $\mathcal{G}_1^m$ to a node in $\mathcal{G}_2^m$), which can only be achieved by assigning a measurement to a line in $\mathcal{L}^m$. In this regard, we define $\mathcal{M}_{\mathcal{L}^m}$ to be the set of line measurements over lines in $\mathcal{L}^m$ and injection measurements over buses incident to $\mathcal{L}^m$, i.e., buses in $\{\mathcal{N}_{1,2}^m\,\cup\,\mathcal{N}_{2,1}^m\}$. 
Based on the measurement assignment rules described in Section~\ref{subsec:Observability}, also summarized in Proposition~\ref{remark:TopologicalObs}, only measurements in $\mathcal{M}_{\mathcal{L}^m}$ could be potentially assigned to a line in $\mathcal{L}^m$. As a result, removing all of the measurements in $\mathcal{M}_{\mathcal{L}^m}$ will guarantee that the system becomes unobservable, as it guarantees that two subgraphs of $\mathcal{G}$ would never be connected by any measurement assignment, making it, thus, impossible to build a spanning tree over $\mathcal{G}$. 
Hence, the removal of $\mathcal{M}_{\mathcal{L}^m}$ is a sufficient condition for causing the unobservability of the system. However, removing the entire set $\mathcal{M}_{\mathcal{L}^m}$ may be more than required to prevent any possible measurement assignment (or constructed tree) from connecting $\mathcal{G}_1^m$ and $\mathcal{G}_2^m$; a goal which could be achieved by the removal of only a subset of $\mathcal{M}_{\mathcal{L}^m}$, as we investigate next.

In this regard, for the pair of subgraphs $\mathcal{G}_1^m$ and $\mathcal{G}_2^m$, resulting from the spanning tree $\mathcal{T}$, we define a set of measurements, $\mathcal{C}^m\in\mathcal{M}_{\mathcal{L}^m}$, for each measurement $m\in\mathcal{M}^A$, to which we refer as a \emph{critical set} of $m$, as follows: 
\begin{definition}\label{def:CriticalSet}
For a measurement $m\in\mathcal{M}^A$ and the pair of subgraphs $\mathcal{G}^m_1(\mathcal{N}^m_1,\mathcal{L}^m_1)$ and $\mathcal{G}^m_2(\mathcal{N}^m_2,\mathcal{L}^m_2)$), a critical set of $m$, denoted by $\mathcal{C}^m\subseteq\mathcal{M}_{\mathcal{L}^m}$, is a maximal set of measurements\footnote{A critical set is defined based on the two subgraphs it aims to reconnect. As these two subgraphs are defined based on the original spanning tree $\mathcal{T}$, $\mathcal{C}^m$ is then dependent on $\mathcal{T}$. However, the notion of a critical set defined here can also be applied to any other pair of adjacent subgraphs. For ease of notation, we do not include $\mathcal{T}$ as an index in the notation of $\mathcal{C}^m$. However, the dependence of $\mathcal{C}^m$ on $\mathcal{T}$, as we have highlighted, is always implied.} within $\mathcal{M}_{\mathcal{L}^m}$ (i.e., the measurements which could be assigned to lines in $\mathcal{L}^m$, following the measurement assignment rules in Proposition~\ref{remark:TopologicalObs}, to connect a bus in $\mathcal{G}^m_1$ to a bus in $\mathcal{G}^m_2$) containing $m$, such that a spanning tree can be formed over each of $\mathcal{G}^m_1(\mathcal{N}^m_1,\mathcal{L}^m_1)$ and $\mathcal{G}^m_2(\mathcal{N}^m_2,\mathcal{L}^m_2)$ using only the measurements in $\{\mathcal{M}^m_1\cup\mathcal{M}^m_2\}\setminus\mathcal{C}^m$.
\end{definition}

In other words, $\mathcal{C}^m$ includes $m$ and a maximal set of redundant measurements in $\mathcal{M}_{\mathcal{L}^m}\setminus\{m\}$. When all of the measurements within $\mathcal{C}^m$ except for an arbitrary one are removed, $\mathcal{G}_1^m$ and $\mathcal{G}_2^m$ can still be connected while having a spanning tree in each, to form a spanning tree over the entire graph $\mathcal{G}$. 
Indeed, when $m$ is removed, any $m'\in\mathcal{C}^m$ can be assigned to a line in $\mathcal{L}^m$ to reconnect $\mathcal{G}_1^m$ and $\mathcal{G}_2^m$, while a spanning tree can still be formed over $\mathcal{G}^m_1$ and $\mathcal{G}^m_2$, using $\{\mathcal{M}^m_1\cup\mathcal{M}^m_2\}\setminus\mathcal{C}^m$.
A critical set, such as $\mathcal{C}^m\in\mathcal{M}_{\mathcal{L}^m}$ is maximal in the sense that if any additional measurement in $\mathcal{M}_{\mathcal{L}^m}$ is added to $\mathcal{C}^m$, no spanning tree could be formed in $\mathcal{G}^m_1(\mathcal{N}^m_1,\mathcal{L}^m_1)$ or $\mathcal{G}^m_2(\mathcal{N}^m_2,\mathcal{L}^m_2)$ using measurements in $\{\mathcal{M}^m_1\cup\mathcal{M}^m_2\}\setminus\mathcal{C}^m$. 
From an algebraic perspective, $\mathcal{C}^m$ is such that the Jacobian matrices of $\mathcal{G}_1^m$ and $\mathcal{G}_2^m$ using only the measurements in $\mathcal{M}^m_1\setminus \{\mathcal{M}^m_1\cap\mathcal{C}^m\}$ and $\mathcal{M}^m_2\setminus \{\mathcal{M}^m_2\cap\,\mathcal{C}^m\}$, denoted by $\boldsymbol{H}^{(-\mathcal{C}^m)}_1$ and $\boldsymbol{H}^{(-\mathcal{C}^m)}_2$, are full column rank, i.e., rank$(\boldsymbol{H}^{(-\mathcal{C}^m)}_1)=N^m_1-1$ and rank$(\boldsymbol{H}^{(-\mathcal{C}^m)}_2)=N^m_2-1$. However, the addition of any additional measurement in $\mathcal{M}_{\mathcal{L}^m}$ to $\mathcal{C}^m$ would lead to rank$(\boldsymbol{H}^{(-\mathcal{C}^m)}_1)<N^m_1-1$ or rank$(\boldsymbol{H}^{(-\mathcal{C}^m)}_2)<N^m_2-1$.
Hence, finding $\mathcal{C}^m$ corresponds to finding a maximal set of redundant measurements in $\mathcal{M}_{\mathcal{L}^m}\setminus\{m\}$. This maximal set may not be unique (as we will demonstrate in the analysis that ensues). Indeed, any set of measurements that meets Definition~\ref{def:CriticalSet}, is a critical set of $m\in\mathcal{M}^A$, considering the pair of subgraphs $\mathcal{G}^m_1(\mathcal{N}^m_1,\mathcal{L}^m_1)$ and $\mathcal{G}^m_2(\mathcal{N}^m_2,\mathcal{L}^m_2)$. Here, in general, a set of measurements is redundant if its removal does not affect the rank of the Jacobian matrix in the studied graph/subgraph or, equivalently, does not prevent the possibility of building a spanning tree over the studied graph/subgraph.
%
%
%
%
%
%
%
We next introduce a set of rules for building a critical set -- which is not necessarily unique -- of a certain measurement, following which we provide a step-by-step procedure for building such critical sets. As such, a critical set can be derived for each of the $N-1$ measurements in $\mathcal{M}^A$, i.e., the measurement assigned as part of the original tree. This set of $N-1$ critical sets is then used, analyzed, and extended for analyzing the observability of the system. 

\begin{figure}[t!]
  \begin{center}
   \vspace{-0.35cm}
    \includegraphics[width=8.5cm]{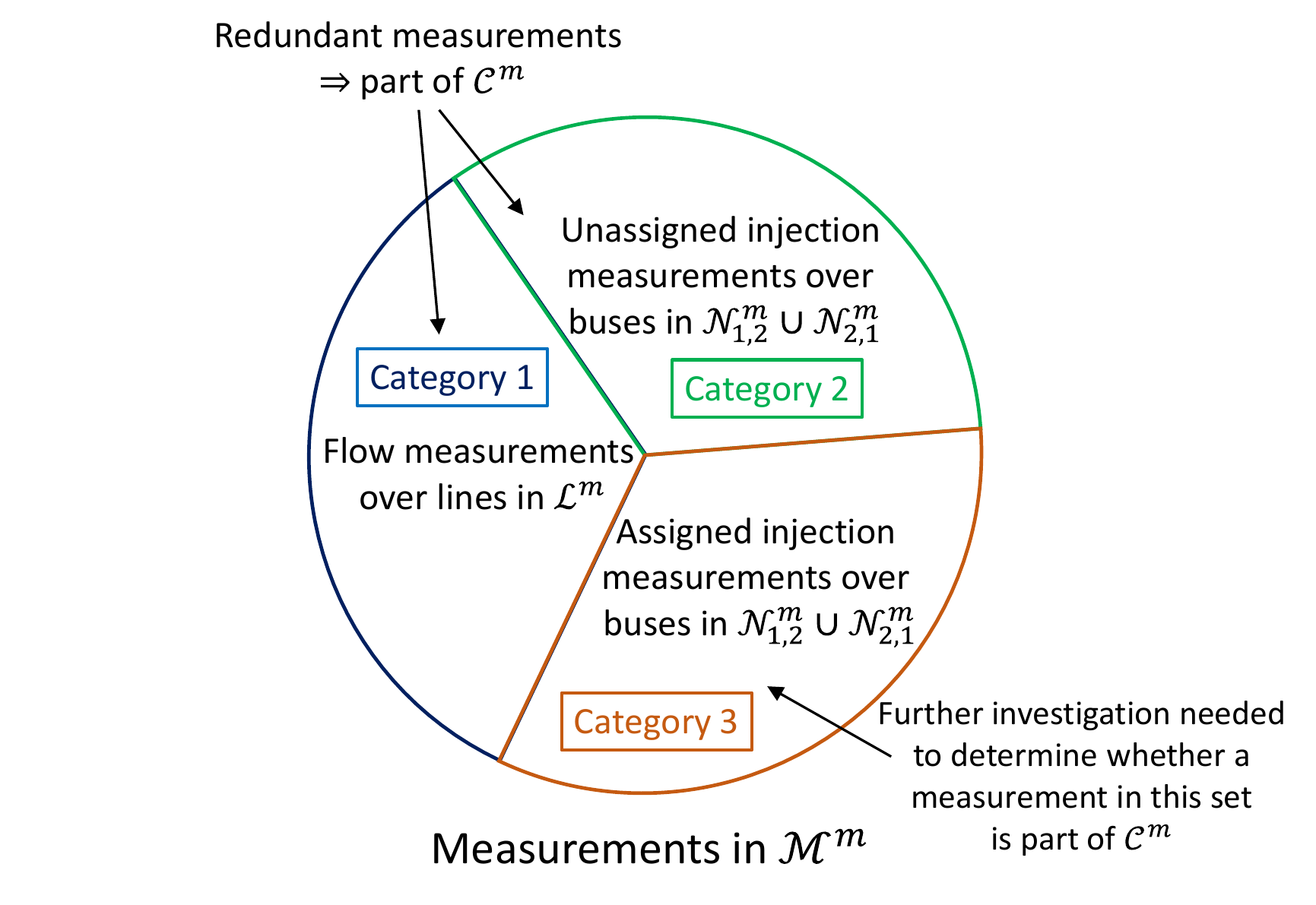}
    \vspace{-0.5cm}
    \caption{\label{fig:MeasurementCategories} The three categories of measurements in $\mathcal{M}^m$.}
  \end{center}\vspace{-0.65cm}
\end{figure}

For a measurement $m$, we let $\mathcal{M}^m=\mathcal{M}_{\mathcal{L}^m}\setminus\{m\}$, and investigate the redundant subset of measurements in $\mathcal{M}^m$ to obtain a critical set $\mathcal{C}^m$. 
Given that $m$ is always considered to be part of $\mathcal{C}^m$, $m$ is always added to $\mathcal{C}^m$ after investigating the measurements in $\mathcal{M}^m$. 
An example of the set $\mathcal{M}^m$ for $m\triangleq F_2$ is shown in Fig.~\ref{fig:IEEE14BusSystemCriticalSetExample}. 
The measurements in $\mathcal{M}^m$ can be grouped into three different categories: 1) flow measurements over lines in $\mathcal{L}^m$, which we denote by $\mathcal{M}^m_F$,\footnote{Such measurements are unassigned measurements. In fact, if $m$ is a line measurement, lines in $\mathcal{L}^m$ would form a loop with $f(m)$ and hence cannot be part of the original spanning tree. Moreover, if $m$ is an injection measurement, $\mathcal{M}^m_F$ would be an empty set since, otherwise, based on the spanning tree building method described in Section~\ref{subsec:Observability} and originally presented in~\cite{TopologicalObs1}, one of the measurements in $\mathcal{M}^m_F$ would have been assigned to a line in $\mathcal{L}^m$, and $m$ would not have been part of $\mathcal{M}^A$. As a result, measurements in $\mathcal{M}^m_F$ are redundant.} 2) unassigned injection measurements over buses in $\mathcal{N}_{1,2}^m \cup \mathcal{N}_{2,1}^m$, and 3) assigned injection measurements over buses in $\mathcal{N}_{1,2}^m \cup \mathcal{N}_{2,1}^m$, i.e., measurements in $\mathcal{M}^m\cap\mathcal{M}^A$. A representation of this partition is shown in Fig.~\ref{fig:MeasurementCategories}

Since the measurements in $\mathcal{M}^m_F$ and the unassigned injection measurements over buses in $\mathcal{N}_{1,2}^m \,\cup\, \mathcal{N}_{2,1}^m$ (which are the first two categories of measurements in $\mathcal{M}^m$, as shown in Fig.~\ref{fig:MeasurementCategories}) are redundant, they are part of $\mathcal{C}^m$. For example, consider the injection measurement, $I_4$, over bus $4$ in Fig.~\ref{fig:IEEE14BusSystemCriticalSetExample}. $I_4$ is in $\mathcal{M}^{F_2}$ and is a redundant measurement since it was not assigned to any line as part of the original tree $\mathcal{T}$. 
Thus, $I_4\in\mathcal{C}^{F_2}$.

Now, when $m'\in\mathcal{M}^m$ is assigned as part of the original assignment function, i.e., $m'\in\mathcal{M}^m\cap\mathcal{M}^A$ (which corresponds to the third category of measurements in $\mathcal{M}^m$, indicated in Fig.~\ref{fig:MeasurementCategories}), then additional investigation is needed to determine whether $m'$ is redundant, and hence, whether it can be considered in $\mathcal{C}^m$. This process will explore all alternative ways of building a spanning tree in the subgraph in which $m'$ is assigned, to determine whether $m'$ is a redundant measurement. 

In this regard, consider that $m'\in\mathcal{M}^m$ is assigned to a line $l'$, i.e., $f(m')=l'\in\mathcal{B}$. If $m'$ is to be reassigned to a line $l\in\mathcal{L}^m$, $\mathcal{T}_1^m$ and $\mathcal{T}_2^m$ will be reconnected, but since $m'$ was originally assigned as part of the original tree, another portion of the tree gets disconnected by this reassignment of $m'$ to $l$ instead of $l'$. Hence, $m'$ would be redundant and can, thus, be part of $\mathcal{C}^m$ if another measurement can be used to reconnect the subgraph which was disconnected by the reassignment of $m'$ from $l'$ to $l$. In other words, $m'$ can be part of $\mathcal{C}^m$ if the measurement set within the subgraph that was disconnected by the reassignment of $m'$ to $l$ instead of $l'$ allows building a spanning tree over this subgraph. This can be assessed by analyzing the redundant measurements within this subgraph to investigate all other possible spanning trees which can be formed, hence, not limiting the analysis to the original spanning tree $\mathcal{T}$. For example, consider the injection measurement over bus $13$ in Fig.~\ref{fig:IEEE14BusSystemCriticalSetExample}, which we denote by $I_{13}$. 
$I_{13}$ has been assigned to line $12$ as part of the original spanning tree. Hence, if $I_{13}$ is assigned to line $20$ to reconnect $\mathcal{T}^{F_2}_1$ and $\mathcal{T}^{F_2}_2$ after measurement $F_2$ is removed, it cannot be assigned to line $12$ anymore which will split $\mathcal{T}^{F_2}_1$ into two subtrees, one formed by buses $\{12,\, 13\}$ and line $19$ and the other subtree composed of buses $\{6,\, 10,\, 1,\, 5\}$ and lines $\{13,\, 10,\, 1\}$. We denote these two subtrees by $\mathcal{T}^{F_2}_{1,1}$ and $\mathcal{T}^{F_2}_{1,2}$, respectively. In this respect, if another measurement can replace $I_{13}$ in reconnecting $\mathcal{T}^{F_2}_{1,1}$ and $\mathcal{T}^{F_2}_{1,2}$, then $I_{13}$ would be redundant and can be assigned to line $20$ and, hence, should be part of $\mathcal{C}^{F_2}$. 
To this end, consider the injection measurement over bus $12$, denoted by $I_{12}$, which was not part of the original spanning tree assignment, i.e., $I_{12}\notin\mathcal{M}^A$. $I_{12}$ can be assigned to line $11$ to reconnect $\mathcal{T}^{F_2}_{1,1}$ and $\mathcal{T}^{F_2}_{1,2}$ in case $I_{13}$ is reassigned to line $20$ instead of line $12$. Hence, $I_{13}$ is indeed redundant, resulting in $I_{13}\in\mathcal{C}^{F_2}$. Thus, the assignment of $I_{12}$ to line $11$ will lead to modifications to $\mathcal{T}^{F_2}_1$. However, here, the goal is not to identify the new form of $\mathcal{T}^{F_2}_1$, but rather to merely investigate whether or not a new form of $\mathcal{T}^{F_2}_1$ can be constructed using the measurement set in $\mathcal{G}^{F_2}_1$. To generalize the analysis in this example, we next provide a general discussion of measurements in $\mathcal{M}^m\cap\mathcal{M}^A$ (i.e., third category in Fig.~\ref{fig:MeasurementCategories}) which allows determining whether a measurement in this set is part of $\mathcal{C}^m$.       




More generally, consider $m'\in\mathcal{M}^m\cap\mathcal{M}^A$ to be a measurement assigned to a branch $l'=f(m')$ in $\mathcal{B}_1^m$, and let $\mathcal{M}_1^m$ be the set of measurements in $\mathcal{G}_1^m$. Reassigning $m'$ to $l\in\mathcal{L}^m$ instead of $l'$, to reconnect $\mathcal{T}_1^m$ and $\mathcal{T}_2^m$, will split $\mathcal{T}_1^m$ into two subtrees $\mathcal{T}^{m}_{1,1}$ and $\mathcal{T}^{m}_{1,2}$. These trees, respectively, span subgraphs $\mathcal{G}^{m}_{1,1}$ and $\mathcal{G}^{m}_{1,2}$. Let $\mathcal{M}^{m}_{1,1}$ and $\mathcal{M}^{m}_{1,2}$ be the sets of measurements in $\mathcal{G}^{m}_{1,1}$ and $\mathcal{G}^{m}_{1,2}$. $m'$ can be reassigned to $l$ only if some measurement in $\mathcal{M}^m_1$ can reconnect $\mathcal{T}^{m}_{1,1}$ and $\mathcal{T}^{m}_{1,2}$. Hence, this corresponds to finding a measurement assignment that connects the two subtrees $\mathcal{T}^m_{1,1}$ and $\mathcal{T}^m_{1,2}$. As discussed in Section~\ref{subsec:Observability}, 
two subtrees can be connected by using a measurement assignment if the processing of an unassigned boundary injection in one of them reaches a node in the other. We denote such unassigned boundary injections by \emph{backup boundary injections}, defined as follows: 
\begin{definition}
A measurement $m''\in\mathcal{M}^{m}_1\setminus\{\mathcal{M}_1^m\cap\mathcal{M}^m\}$ is a \emph{backup boundary injection} for a measurement $m'$, if $m''$ is a boundary injection which can reconnect -- following the boundary injection assignment procedure in~\cite{TopologicalObs1} -- the subtrees $\mathcal{T}^{m}_{1,1}$ and $\mathcal{T}^{m}_{1,2}$ generated by the reassignment of $m'\in\mathcal{M}^m\cap\mathcal{M}^A$ to a line $l\in\mathcal{L}^m$ instead of its original line assignment $f(m')=l'$. The set of all such backup boundary injections for this measurement $m'\in\mathcal{M}^m\cap\mathcal{M}^A$ is referred to as the \emph{backup boundary injection set} of $m'$ and is denoted by $\mathcal{I}_{b-m'}^{m}$.  
\end{definition}

Since the tree building algorithm in~\cite{TopologicalObs1} is based on connecting subtrees (i.e. tree components) -- to build a full spanning tree -- by starting from an unassigned boundary injection in a certain subtree (as a source) to reach a node in another subtree (as a sink), this algorithm can be readily employed to identify a backup boundary injection for a measurement $m'\in\mathcal{M}^m\cap\mathcal{M}^A$. To this end, to find a backup boundary injection of $m'$, we run the algorithm in~\cite{TopologicalObs1} by starting from an unassigned boundary injection in either $\mathcal{T}^m_{1,1}$ or $\mathcal{T}^m_{1,2}$ and checking whether the algorithm reaches a bus in $\mathcal{T}^m_{1,2}$ or $\mathcal{T}^m_{1,1}$, respectively. As such, using the spanning tree building algorithm provided in~\cite[Fig. 1]{TopologicalObs1}, one can identify the backup boundary injections for each injection measurement in $m'\in\mathcal{M}^m\cap\mathcal{M}^A$. Here, we note that, following our proposed method for constructing $\mathcal{C}^m$, a backup boundary injection cannot be an injection measurement in $\mathcal{M}^m$, since if $m''\in\mathcal{M}^m$ and $m''$ is unassigned, $m''$ will itself be part of $\mathcal{C}^m$, as previously discussed. 

\begin{figure}[t!]
  \begin{center}
   \vspace{-0.35cm}
    \includegraphics[width=4.2cm]{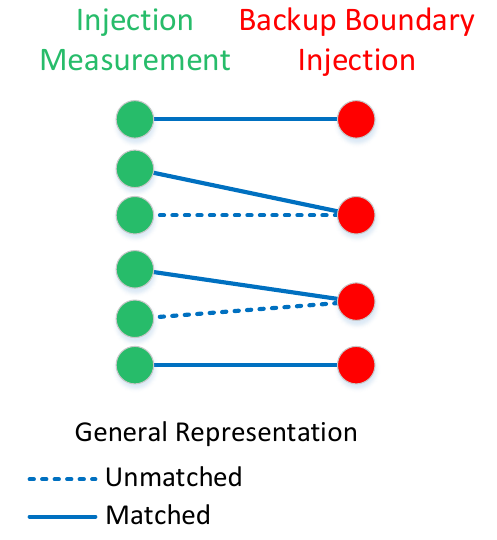}
    \vspace{-0.4cm}
    \caption{\label{fig:MaxMatchingBackupBoundary} Maximum matching over the ``injection measurements - backup boundary injections bipartite graph''.}
  \end{center}\vspace{-0.95cm}
\end{figure}

Therefore, an assigned injection measurement $m'\in\mathcal{M}^m\cap\mathcal{M}^A$ is a redundant measurement and is, as a result, part of $\mathcal{C}^m$ if it has a nonempty boundary injection set $\mathcal{I}_{b-m'}^m$. However, a boundary injection may be part of multiple backup boundary injection sets. In this regard, based on the measurement assignment rules, an injection measurement can be assigned to only one line at a time. Hence, an unassigned boundary injection can act as a backup boundary injection for \emph{only one} measurement in $\mathcal{M}^m\cap\mathcal{M}^A$, at a time. Thus, if two measurements $m_1$ and $m_2$ in $\mathcal{M}^m\cap\mathcal{M}^A$ have only one and the same backup boundary injection, only one of them can be concurrently in $\mathcal{C}^m$, which indicates that a critical set $\mathcal{C}^m$ of $m$ may not be unique. As a result, due to this one-to-one assignment requirement between backup boundary injections and injection measurements in $\mathcal{M}^m\cap\mathcal{M}^A$, finding this assignment can be performed by solving a maximum matching problem over a bipartite graph\footnote{A matching over a graph is a subset of edges sharing no vertices. A maximum matching is a matching having the maximum possible number of edges~\cite{GraphTheoryWest}.}, as the one shown in Fig.~\ref{fig:MaxMatchingBackupBoundary}. We refer to this graph as the \emph{injection measurements - backup boundary injections bipartite graph}.

In this bipartite graph, the left-side nodes denote the injection measurements in $\mathcal{M}^m\cap\mathcal{M}^A$ and right-side nodes denote the union of their backup boundary injections, $\bigcup\limits_{m'\in\mathcal{M}^m\cap\mathcal{M}^A}\mathcal{I}_{b-m'}^m$, in which each node represents one backup boundary injection. In this bipartite graph, an edge exists between a node $m'\in\mathcal{M}^m\cap\mathcal{M}^A$, on the left-side of the graph, and a boundary injection $m''$, on the right-side of the graph, if $m''\in\mathcal{I}^m_{b-m'}$. Here, we note that a boundary injection can be simultaneously part of different backup boundary injection sets. Hence, finding the injection measurements in $\mathcal{M}^m\cap\mathcal{M}^A$ which are part of a critical set of $m$, $\mathcal{C}^m$, requires solving a maximum matching problem over this bipartite graph\footnote{The solution of a maximum matching problem over a bipartite graph can be efficiently obtained in polynomial time by transforming the matching problem into a max-flow problem, which can be solved in polynomial time using various known algorithms such as \emph{Ford-Fulkerson}~\cite{GraphTheoryWest}.}.  
%
As a result, the matched left-side nodes in a maximum matching over the ``injection measurements - backup boundary injections bipartite graph'' are the injection measurements in $\mathcal{M}^m\cap\mathcal{M}^A$ which will be part of $\mathcal{C}^m$. Here, we note that this leads to the identification of a critical set $\mathcal{C}^m$ of $m$, as $\mathcal{C}^m$ may not be unique given that a maximum matching over a bipartite graph may not be unique. Indeed, each possible maximum matching would lead to a different $\mathcal{C}^m$. 

Based on these introduced rules for building critical sets, Algorithm~\ref{alg:CriticalSetAlgorithm} provides a structured step-by-step method for building a critical set, $\mathcal{C}^m$, for each measurement $m\in\mathcal{M}^A$. \vspace{-0.1cm}

\begin{algorithm}[!t]
\caption{Critical sets step-by-step procedure}
\label{alg:CriticalSetAlgorithm}
{\footnotesize \begin{algorithmic}[1]
\REQUIRE Power system 1-line diagram $\mathcal{G}(\mathcal{N},\mathcal{L})$, measurement set $\mathcal{M}$, spanning tree $\mathcal{T}(\mathcal{N},\mathcal{B})$, set of assigned measurements $\mathcal{M}^A$, assignment function $f(.)$$:\mathcal{M}^A\rightarrow \mathcal{B}$  
\ENSURE Critical set $\mathcal{C}^m$ for all measurements $m\in\mathcal{M}^A$
\FOR {$m\in\mathcal{M}^A$} 
\STATE Characterize $\mathcal{T}_1^m$, $\mathcal{T}_2^m$, $\mathcal{N}_{1,2}^m$, $\mathcal{N}_{2,1}^m$, $\mathcal{L}^m$, $\mathcal{M}^m$
\STATE Initialize $\mathcal{C}^m$
\STATE Initialize $\mathcal{M}^m_{\textrm{test}}$
\STATE Add $m$ to $\mathcal{C}^m$
\FOR{$m'\in\mathcal{M}^m$}
\IF {$m'$ is a flow measurement}
\STATE Add $m'$ to $\mathcal{C}^m$
\ENDIF
\IF {$m'$ is an injection measurement}
\IF{$m'\notin\mathcal{M}^A$}
\STATE Add $m'$ to $\mathcal{C}^m$
\ENDIF
\IF{$m'\in\mathcal{M}^A$}
\STATE Characterize its backup boundary injection set $\mathcal{I}^m_{b-m'}$
\STATE Add $m'$ to $\mathcal{M}^m_{\textrm{test}}$
\ENDIF
\ENDIF
\ENDFOR
\STATE Solve maximum matching over the ``injection measurements - backup boundary injections bipartite graph''
\FOR {$m'\in\mathcal{M}^m_{\textrm{test}}$}
\IF {$m'$ is a matched node as part of the maximum matching}
\STATE Add $m'$ to $\mathcal{C}^m$
\ENDIF
\ENDFOR
\ENDFOR
\RETURN Critical set $\mathcal{C}^m$ for all measurements $m\in\mathcal{M}^A$
\end{algorithmic}}
\end{algorithm}

\subsection{Example and Case Analysis}
As an example of characterizing critical sets of the various assigned measurements, as part of a spanning tree, in a power system, we consider the IEEE 14-bus system in Fig.~\ref{fig:IEEE14Bus}. In this example, we denote an injection measurement over bus $k$ by $I_k$ and a flow measurement over line $k$ by $F_k$. We first consider the measurement over line $2$, $F_2$, for which we find a critical set $\mathcal{C}^{F_2}$ using Algorithm~\ref{alg:CriticalSetAlgorithm}. 

From Fig.~\ref{fig:IEEE14BusSystemCriticalSetExample}, we can see that removing $F_2$ will result in splitting the original spanning tree into two trees, $\mathcal{T}_1^{F_2}(\mathcal{N}^{F_2}_1,\mathcal{B}^{F_2}_1)$ and $\mathcal{T}^{F_2}_2(\mathcal{N}^{F_2}_2,\mathcal{B}^{F_2}_2)$, such that $\mathcal{N}^{F_2}_1=\{1,5,6,10,12,13\}$ and $\mathcal{N}^{F_2}_2=\{2,3,4,7,8,9,11,14\}$ are the sets of nodes of the two trees and  
$\mathcal{B}^{F_2}_1=\{1,10,13,12,19\}$ and $\mathcal{B}^{F_2}_2=\{4,6,8,15,9,16,17\}$ are their sets of edges. These two trees span, respectively, $\mathcal{G}^{F_2}_1$ and $\mathcal{G}^{F_2}_2$, as shown in Fig.~\ref{fig:IEEE14BusSystemCriticalSetExample}. In addition, $\mathcal{N}_{1,2}^{F_2}=\{1,5,10,13\}$, $\mathcal{N}_{2,1}^{F_2}=\{2,4,11,14\}$, $\mathcal{L}^{F_2}=\{2,3,7,18,20\}$, and $\mathcal{M}^{F_2}=\{F_2,\, I_4,\, I_{11},\, I_2, I_1,\, I_5,\, I_{13}\}$. Now, for characterizing a critical set of $F_2$, we explore set $\mathcal{M}^{F_2}$. 

The first measurement in $\mathcal{M}^{F_2}$ is $F_2$. $F_2$ is a flow measurement. 
Hence, $F_2\in\mathcal{C}^{F_2}$. 
The second and third measurements in $\mathcal{M}^{F_2}$ are $I_4$ and $I_{11}$. $I_4$ and $I_{11}$ are unassigned injection measurements, i.e. $I_4\notin\mathcal{M}^A$ and $I_{11}\notin\mathcal{M}^A$. Hence, $\{I_4, I_{11}\}\subseteq\mathcal{C}^{F_2}$. Indeed, $I_4$ can be assigned to line $7$ to reconnect $\mathcal{T}_1^{F_2}$ and $\mathcal{T}_2^{F_2}$, while $I_{11}$ can be assigned to line $18$ for that purpose.
$I_2$ is the fourth measurement in $\mathcal{M}^{F_2}$ and the last remaining injection measurement on $\mathcal{N}^m_{2,1}$ to be explored. $I_2$ is an assigned measurement, originally assigned to line $4$ as part of the spanning tree $\mathcal{T}$. As $I_2\in\mathcal{M}^{A}$, assigning $I_2$ to lines $2$ or $3$ to reconnect $\mathcal{T}_1^{F_2}$ and $\mathcal{T}_2^{F_2}$ will disconnect bus $2$ from the rest of $\mathcal{T}_2^{F_2}$. Hence, we next characterize the backup boundary injection set of $I_2$, i.e. $\mathcal{I}^{F_2}_{b-I_2}$. The only unassigned boundary injection in $\mathcal{G}_2^{F_2}$ that is not part of $\mathcal{M}^{F_2}$ is $I_7$. However, using the algorithm in~\cite[Fig. 1]{TopologicalObs1}, we can observe that starting from $I_7$, the algorithm does not reach bus $2$. Hence, bus $2$ cannot be reconnected to the rest of $\mathcal{T}^{F_2}_2$ using any unassigned boundary injections over buses in $\mathcal{G}_2^{F_2}$. Hence, $\mathcal{I}^{F_2}_{b-I_2}=\emptyset$, and as a result $F_2\notin\mathcal{C}^{F_2}$.   
Similarly, exploring $I_{1}$ and $I_{5}$ -- the fifth and sixth measurements in $\mathcal{M}^{F_2}$ -- which are both assigned measurements, i.e. $\{I_1,\, I_5\}\subseteq\mathcal{M}^A$, shows that they both have empty backup boundary injection sets\footnote{If $I_1$ or $I_5$ are to be reassigned to lines $2$ or $3$, respectively, to reconnect $\mathcal{T}_1^{F_2}$ and $\mathcal{T}_2^{F_2}$, each of these reassignments will split $\mathcal{T}_1^{F_2}$ into two subtrees which cannot be reconnected using the unassigned boundary injection $I_{12}$, as can be shown by a run of the algorithm in~\cite[Fig. 1]{TopologicalObs1}. Here, we note that $I_{12}$ is the only unassigned boundary injection in $\mathcal{G}^{F_2}_1$.}, i.e. $\mathcal{I}^{F_2}_{b-I_1}=\emptyset$ and $\mathcal{I}^{F_2}_{b-I_5}=\emptyset$. Hence, neither $I_1$ nor $I_5$ are part of $\mathcal{C}^{F_2}$. 
The only remaining measurement in $\mathcal{M}^{F_2}$ is $I_{13}$. $I_{13}$ is an assigned measurement, $I_{13}\in\mathcal{M}^A$. As previously discussed in Section~\ref{subsec:CriticalMeas}, when $I_{13}$ is reassigned to line $20$ to reconnect $\mathcal{T}^{F_2}_1$ and $\mathcal{T}^{F_2}_2$, the subtree containing buses $12$ and $13$ and line $19$ gets disconnected from the rest of $\mathcal{T}^{F_2}_1$. Hence, we next characterize the backup boundary injection set of $I_{13}$, i.e. $\mathcal{I}^{F_2}_{b-I_{13}}$. To this end, $I_{12}$, the only unassigned boundary injection measurement in $\mathcal{G}_1^{F_2}$, can be assigned to line $11$ to reconnect the two subtrees, and is the only boundary injection which can do so. Hence, $\mathcal{I}^{F_2}_{b-I_{13}}=\{I_{12}\}$. 
As a result, the ``injection measurements - backup boundary injections bipartite graph'' is composed of only $I_{13}$ on the left-side connected to $\mathcal{I}^{F_2}_{b-I_{13}}=\{I_{12}\}$ on the right-side. Hence, $I_{13}$ is matched to the backup boundary injection $I_{12}$. As a result, $I_{13}\in\mathcal{C}^{F_2}$. 
The processing of $\mathcal{M}^{F_2}$ is thus complete, resulting in $\mathcal{C}^{F_2}=\{F_2,I_{4}, I_{11},I_{13}\}$.

Similarly, Algorithm~\ref{alg:CriticalSetAlgorithm} can be carried out to characterize critical sets of all of the measurements in $\mathcal{M}^A$ in the IEEE 14-bus system in Fig.\ref{fig:IEEE14Bus}. The results are listed in Table~\ref{Tab:CriticalSets}. 

\begin{table}[t!]
	\caption{Critical sets of the measurements in $\mathcal{M}^A$.}\label{Tab:CriticalSets}\vspace{-0.5cm}
	\begin{center}
		\begin{tabular}{|c|c|}
		\hline
			   \textbf{Measurement $(m\in\mathcal{M}^A)$} & \textbf{Critical Set $(\mathcal{C}^m)$}   \\ \hline
                   $F_2$&$\{F_2, I_4, I_{11}, I_{13}\}$ \\ \hline
                   $F_8$ & $\{F_8,I_4,I_7,I_9\}$\\ \hline
                    $F_9$ & $\{F_9,I_4,I_7,I_{11},I_{13}\}$ \\ \hline 
                    $ F_{15}$ & $\{F_{15},I_7\}$  \\ \hline  
                    $I_1$ & $\{I_1,I_4,I_{11},I_{13}\}$ \\ \hline                  
                    $I_2$& $\{I_2,I_4, I_{11}, I_{13}\}$  \\ \hline    
                    $I_3$ & $\{I_3,I_2,I_4\}$\\ \hline 
                    $I_5$ & $\{I_5,I_{11},I_{13}\}$\\ \hline
                    $I_6$ & $\{I_6,I_{11}\}$\\ \hline    
                    $I_9$ & $\{I_9,I_{11}\}$\\ \hline   
                    $I_{13}$ & $\{I_6,I_{12},I_{13}\}$\\ \hline
                    $F_{17}$ & $\{F_{17},I_{9}, I_{13}\}$\\ \hline
                    $F_{19}$ & $\{F_{19},I_6,I_{12}\}$ \\ \hline                       
		\end{tabular}
	\end{center}\vspace{-0.7cm}
\end{table}  

We next discuss the value of critical sets with regard to understanding and analyzing observability attacks. We also introduce the concept of observability sets, a generalization of critical sets, which provides a holistic modeling of observability attacks. 

\textbf{Notation:} We use the following notation in the derivations that ensue. For the Jacobian matrix $\boldsymbol{H}$, we let $\boldsymbol{H}^{(-\mathcal{K}) + (\mathcal{K}')}$ correspond to $\boldsymbol{H}$ but with the removal of the rows corresponding to measurements in $\mathcal{K}$ and the addition of rows corresponding to measurements in $\mathcal{K}'$. \vspace{-0.2cm}    
%
\subsection{Critical Sets and Observability}
Next, we show that the derived critical sets are indispensable for modeling observability attacks. In this regard, we show that the removal of a critical set renders the system unobservable. The proof is carried out using two approaches: 1) A graph-theoretic approach presented in Theorem~\ref{prop:Obs} and showing that no spanning tree could be build over $\mathcal{G}$ when a critical set is removed, and 2) an approach based on linear algebra, presented in Theorem~\ref{prop:Minus1}, where we will prove that the rank of the Jacobian matrix of the power system graph $\mathcal{G}$, when not considering the measurements in a critical set (i.e. after the removal of a critical set), is strictly less than $N-1$ (we, in fact, prove that the rank of the Jacobian matrix becomes $N-2$), making the Jacobian matrix not full column rank and leading the system to become unobservable. These two approaches are presented next.

\emph{1) Proof by Graph-Theoretic Arguments:} 

\begin{theorem}\label{prop:Obs}
Removing a critical set, $\mathcal{C}^m$, renders the system unobservable.  
\end{theorem}
\begin{IEEEproof}
By topological observability, we know that a system is observable if and only if a spanning tree could be formed using an assignment function. Hence, given that $\mathcal{G}^m_1$ and $\mathcal{G}^m_2$ are two disjoint subgraphs of $\mathcal{G}$, any spanning tree over $\mathcal{G}$ must at least assign one measurement to a line in $\mathcal{L}^m$. Otherwise, this tree would not span the whole system.  

When $m$ is removed, the original spanning tree $\mathcal{T}$ is split into two disjoint trees $\mathcal{T}^m_1$ and $\mathcal{T}^m_2$, spanning subgraphs $\mathcal{G}^m_1$ and $\mathcal{G}_2^m$, respectively. This makes each of $\mathcal{G}^m_1$ and $\mathcal{G}^m_2$ observable. However, it does not make the whole system observable as there is a need to connect $\mathcal{G}^m_1$ and $\mathcal{G}_2^m$.
By definition of $\mathcal{C}^m$, removing $\mathcal{C}^m$ makes it impossible to connect $\mathcal{G}^m_1$ and $\mathcal{G}_2^m$, by any measurement assignment, as $\mathcal{C}^m$ contains all measurements which could be assigned to a line in $\mathcal{L}^m$ without losing the observability within one of the two subgraphs. In fact, if $\mathcal{C}^m$ is removed, based on Proposition~\ref{remark:TopologicalObs}, for any measurement to be assignable to a line in $\mathcal{L}^m$ it must be in $\mathcal{M}_{\mathcal{L}^m}\setminus\mathcal{C}^m$, in order to either connect $\mathcal{T}^m_1$ and $\mathcal{T}^m_2$ or interconnect disjoint components of $\mathcal{T}^m_1$ and $\mathcal{T}^m_2$ (or disjoint tree components within each of $\mathcal{G}^m_1$ and $\mathcal{G}^m_2$) to span the entire $\mathcal{G}$. However, none of these measurements in $\mathcal{M}_{\mathcal{L}^m}\setminus\mathcal{C}^m$ can be assigned to a line in $\mathcal{L}^m$, as none of these measurements is redundant. In fact, if any of these measurements in $\mathcal{M}_{\mathcal{L}^m}\setminus\mathcal{C}^m$ were redundant, this measurement would have been part of $\mathcal{C}^m$, as otherwise, it would have contradicted the maximality property of $\mathcal{C}^m$. As such, if $\mathcal{C}^m$ is removed, $\mathcal{G}$ would contain no spanning tree that could be built according to the rules of Proposition~\ref{remark:TopologicalObs}, which makes the system unobservable.
\end{IEEEproof}
%
%

\emph{2) Proof by Linear Algebra:}\newline\indent
In addition, we prove in the next theorem (Theorem~\ref{prop:Minus1}) that removing a full critical set decreases the rank of the Jacobian matrix by 1, which as a result makes the system unobservable. However, we first present the following preliminary lemma, which is essential for the proof of Theorem~\ref{prop:Minus1}. 
\begin{lemma}\label{lemma:HypotheticalFlow}
Let $m\in\mathcal{M}$ be an injection measurement over a bus $\eta$ that is assigned to a line $l$, $f(m)=l$. Then, replacing $m$ by a hypothetical line flow measurement $m'$ over line $l$ will not affect the rank of matrix $\boldsymbol{H}$. In other words, let $\boldsymbol{H}^{(-m)+(m')}$ be the the Jacobian matrix with the removal of the row corresponding to measurement $m$ and the addition of the row corresponding to the hypothetical measurement $m'$, then rank$(\boldsymbol{H})=$ rank$(\boldsymbol{H}^{(-m)+(m')})$.   
\end{lemma}
\begin{IEEEproof}
Since $m$ is assigned, i.e. is part of an original spanning tree measurement assignment, removing it will split the original spanning tree into two subtrees $\mathcal{T}^m_1$ and $\mathcal{T}^m_2$. 
If $m'$ existed, $m'$ can reconnect $\mathcal{T}^m_1$ and $\mathcal{T}^m_2$, as it is a measurement over line $l$, which is part of $\mathcal{L}^m$. Hence, replacing $m$ by $m'$ will not affect the connectivity of the spanning tree and, hence, rank$(\boldsymbol{H})=$ rank$(\boldsymbol{H}^{(-m)+(m')})$.   
\end{IEEEproof}

\begin{theorem}\label{prop:Minus1}
For $m\in\mathcal{M}^A$, removing $\mathcal{C}^m$ results in rank$(\boldsymbol{H}^{(-\mathcal{C}^m)})$ $=$ rank$(\boldsymbol{H})-1$, which makes the system unobservable.
\end{theorem}
\begin{IEEEproof}
Since the system is originally fully observable, rank$(\boldsymbol{H})=N-1$. Now, let $\mathcal{M}^m_1$ and $\mathcal{M}^m_2$ be the measurement sets in subgraphs $\mathcal{G}^m_1$ and $\mathcal{G}^m_2$, respectively, and let $\boldsymbol{H}^{(-\mathcal{C}^m)}_1$ and $\boldsymbol{H}^{(-\mathcal{C}^m)}_2$ be the Jacobian matrices of $\mathcal{G}^m_1$ and $\mathcal{G}^m_2$, respectively, composed of measurements in $\mathcal{M}^m_1\setminus \{\mathcal{M}^m_1\cap\mathcal{C}^m\}$ and $\mathcal{M}^m_2\setminus \{\mathcal{M}^m_2\cap\,\mathcal{C}^m\}$. 
Since $\mathcal{T}^m_1$ and $\mathcal{T}^m_2$ respectively span $\mathcal{G}_1^m$ and $\mathcal{G}_2^m$, this implies that rank$(\boldsymbol{H}^{(-\mathcal{C}^m)}_1)=N^m_1-1$ and rank$(\boldsymbol{H}^{(-\mathcal{C}^m)}_2)=N^m_2-1$. In addition, let $m'\in\mathcal{M}_1^m\setminus \{\mathcal{M}_1^m\cap\mathcal{C}^m\}$ be an injection measurement over a bus in $\mathcal{N}_{1,2}^m$. Since $m'\in\mathcal{M}_1^m\setminus \{\mathcal{M}^m_1\cap\mathcal{C}^m\}$, then $m'$ is assigned to a certain branch $b'=f(m')\in\mathcal{T}^m_1$; otherwise, $m'$ would have also been in $\mathcal{C}^m$. By Lemma~\ref{lemma:HypotheticalFlow}, $m'$ can be replaced by a hypothetical line flow measurement over $b'$ without affecting the rank of $\boldsymbol{H}^{(-\mathcal{C}^m)}_1$. As such, let $\boldsymbol{H}^{(-\mathcal{C}^m)'}_1$ be the same as $\boldsymbol{H}^{(-\mathcal{C}^m)}_1$ but replacing any row corresponding to an injection measurement in $\mathcal{N}_{1,2}^m$ by its corresponding hypothetical line flow measurement. The same can be done to form Jacobian matrix $\boldsymbol{H}^{(-\mathcal{C}^m)'}_2$ from $\boldsymbol{H}^{(-\mathcal{C}^m)}_2$. By Lemma~\ref{lemma:HypotheticalFlow}, rank$(\boldsymbol{H}^{(-\mathcal{C}^m)'}_1)$$=$rank$(\boldsymbol{H}^{(-\mathcal{C}^m)}_1)=N^m_1-1$ and rank$(\boldsymbol{H}^{(-\mathcal{C}^m)'}_2)$$=$rank$(\boldsymbol{H}^{(-\mathcal{C}^m)}_2)=N^m_2-1$.  

Now, let us return to $\boldsymbol{H}^{(-\mathcal{C}^m)}$. By rearranging its elements to include first the measurements in $\mathcal{M}_1^m\setminus \{\mathcal{M}_1^m\cap\mathcal{C}^m\}$ then the elements of $\mathcal{M}_2^m\setminus \{\mathcal{M}_2^m\cap\mathcal{C}^m$\}, $\boldsymbol{H}^{(-\mathcal{C}^m)}$ can be written as {\small $\boldsymbol{H}^{(-\mathcal{C}^m)}=\left[ {\begin{array}{c} \boldsymbol{H}^{(-\mathcal{C}^m)}_1 \\\boldsymbol{H}^{(-\mathcal{C}^m)}_2 \end{array}} \right]$}. In this respect, 
{\small\begin{flalign}\nonumber
\textrm{rank}(\boldsymbol{H}^{(-\mathcal{C}^m)})&=\textrm{rank}\Big(\left[ {\begin{array}{c} \boldsymbol{H}^{(-\mathcal{C}^m)}_1 \\\boldsymbol{H}^{(-\mathcal{C}^m)}_2 \end{array}} \right]\Big)&\nonumber\\
&=\textrm{rank}\Big(\left[ {\begin{array}{cc} \boldsymbol{H}^{(-\mathcal{C}^m)'}_1 & \boldsymbol{0}\\ \boldsymbol{0} &\boldsymbol{H}^{(-\mathcal{C}^m)'}_2 \end{array}} \right]\Big)&\nonumber\\
&=(N^m_1-1)+(N^m_2-1)&\nonumber\\
&=N-2=\textrm{rank}(\boldsymbol{H})-1.& \nonumber                      
\end{flalign}}
Therefore, rank$(\boldsymbol{H}^{(-\mathcal{C}^m)})=N-2<N-1 \Rightarrow \boldsymbol{H}^{(-\mathcal{C}^m)}$ is not full column rank $\Rightarrow$ the system becomes unobservable after the removal of any critical set $\mathcal{C}^m$.
\end{IEEEproof}

\emph{3) Analysis and Discussion:}

Theorem~\ref{prop:Minus1} shows the effect of the removal of a single critical set on the rank of the Jacobian matrix. Theorem~\ref{prop:Obs} and Theorem~\ref{prop:Minus1} provide key conditions for observability of the power system under observability attacks. In fact, the contrapositives of these theorems show that if a power system is fully observable, then the investigated observability attack (i.e. the removal of measurements) did not result in removing a full critical set. 
In addition, using Theorem~\ref{prop:Obs} and Theorem~\ref{prop:Minus1}, we can also show that a critical set $\mathcal{C}^m$ is a minimal set of measurements including $m$ which when removed renders the system unobservable. In fact, as shown in Theorem~\ref{prop:Obs} and Theorem~\ref{prop:Minus1}, removing $\mathcal{C}^m$ renders the system unobservable while, if all measurements in $\mathcal{C}^m$ are removed except for one arbitrary measurement $m'$, the system would still remain observable. In fact, by Definition~\ref{def:CriticalSet}, this $m'$ can be assigned to a line $l\in\mathcal{L}^m$ to connect $\mathcal{G}^m_1$ and $\mathcal{G}^m_2$, while having a spanning tree over each of $\mathcal{G}^m_1$ and $\mathcal{G}^m_2$, which leads to a spanning tree over $\mathcal{G}(\mathcal{N},\mathcal{L})$ and makes the system observable.

Moreover, based on Theorem~\ref{prop:Obs} and Theorem~\ref{prop:Minus1}, the critical measurements\footnote{In power systems, a critical measurement is a single measurement which when removed renders the system unobservable~\cite{Abur}.} of a power system can be characterized using the notion of critical sets, as shown in the following corollary. 
\begin{corollary}\label{cor:CriticalMeas}
$m$ is a critical measurement if and only if its only critical set is $\mathcal{C}^m=\{m\}$. 
\end{corollary}
\begin{IEEEproof}
By definition, if $m$ is a critical measurement, removing it will render the system unobservable. Hence, if the critical set of $m$ is such that $\mathcal{C}^m\supset m$, then removing $m$ would not affect the observabilty of the system since any other measurement $m'\in\mathcal{C}^m\setminus\{m\}$ can be used to replace $m$ and reconnect the tree. As such, $\mathcal{C}^m\supset \{m\}$ $\Rightarrow$ $m$ is not a critical measurement, which proves the contrapositive: $m$ is critical $\Rightarrow$ $m$ is the only element in its critical set, i.e. $\{m\}=\mathcal{C}^m$. Conversely, if $m$ is the only element in its critical set, its removal constitutes removing a complete critical set, which by Theorem~\ref{prop:Obs} and Theorem~\ref{prop:Minus1}, renders the system unobservable. As a result, $\mathcal{C}^m=\{m\}$ $\Rightarrow$ $m$ is a critical measurement. 
Thus, $m$ is critical if and only if $\mathcal{C}^m=\{m\}$. 
\end{IEEEproof}

Next, we extend this concept to account for the interconnection between multiple critical sets. 

\subsection{Observability Sets}
For a measurement $m'$ to be in a critical set of a measurement $m$, i.e. $m'\in\mathcal{C}^m$, a critical set of $m'$, $\mathcal{C}^{m'}$, must contain measurements other than $m'$, i.e. $\mathcal{C}^{m'}\supset \{m'\}$. Otherwise, $m'$ would not be redundant. 
For example, consider injection measurements $I_6$ and $I_9$. Removing $I_6$ and $I_9$ will render the system unobservable -- even though $I_6$ and $I_9$ do not form a critical set -- since $\mathcal{C}^{I_6}=\{I_6,I_{11}\}$ and $\mathcal{C}^{I_9}=\{I_9,I_{11}\}$. As such, if $I_6$ is removed, $I_{11}$ can be used to replace $I_6$ since $I_{11}\in\mathcal{C}^{I_6}$. However, if $I_9$ is also removed, even though $I_{11}\in\mathcal{C}^{I_9}$, $I_{11}$ cannot be used to replace $I_9$ since $I_{11}$ has already been used as a replacement to $I_6$. Therefore, removing $I_9$ and $I_6$ does render the system unobservable. Indeed, rank$(\boldsymbol{H}^{-(I_6)-(I_9)})=12<N-1=13$. 
This concept can be extended to the interconnection between multiple critical sets. For example, consider $F_2$, $I_1$, $I_2$, and $I_5$ and their critical sets shown in Table~\ref{Tab:CriticalSets}. We can see that $F_2$, $I_1$, and $I_2$ have critical sets sharing measurements $I_4$, $I_{11}$, and $I_{13}$. Hence, if $F_2$, $I_1$, and $I_2$ are removed, $I_4$, $I_{11}$, and $I_{13}$ are assigned, one to each of these measurements, to preserve system observability and, hence, cannot be used as part of further critical sets in case further measurements are removed. Hence, since $\mathcal{C}^{I_5}=\{I_5, I_{11}, I_{13}\}$, removing $F_2$, $I_1$, $I_2$ and $I_5$ will render the system unobservable, even though $\{F_2,\, I_1,\, I_2,\, I_5\}$ is not a critical set. 

In this respect, the concept of critical sets must be further developed to yield a general graph-theoretic concept of observability attacks. 
To this end, we define a graph, dubbed \emph{``critical sets - system measurements bipartite graph''}, as follows:
\begin{definition} 
A \emph{critical sets - system measurements bipartite graph} (CS-SMBG) is a bipartite graph in which each left-hand side node represents one of the critical sets of the power system (such that the set of left-hand side nodes include only one critical set $\mathcal{C}^m$ for each $m\in\mathcal{M}^A$), and the right-hand side nodes represent the entire measurement set $\mathcal{M}$ such that each right-hand side node represents one measurement in the measurement set $\mathcal{M}$. In this respect, an edge between a critical set $\mathcal{C}^i$ and a measurement $j$ exists if $j\in\mathcal{C}^i$. 
\end{definition}

An example of this bipartite graph is shown in Fig.~\ref{fig:CriticalSetSystmMeasBipartitesDIA}. Here, we note that since a critical set $\mathcal{C}^m$ might not be unique, different versions of a CS-SMBG can be constructed for a single power system, depending on the choice of a critical set for each $m\in\mathcal{M}^A$, which depends on the choice of the assigned measurements $\mathcal{M}^A$ (associated with the original spanning tree $\mathcal{T}$). Based on this formulation, a general concept of observability is established in Theorem~\ref{theorem:ObservabilityGeneral}. 

\begin{theorem}\label{theorem:ObservabilityGeneral}
If the system is observable, then any maximum matching over any CS-SMBG must include all of the critical sets of this CS-SMBG.   
\end{theorem}   
\begin{IEEEproof}
We prove this theorem by proving its contrapositive which is the following: if a maximum matching does not include all of the critical sets of a certain CS-SMBG (i.e. at least one critical set is unmatched in a maximum matching over this CS-SMBG), then the system is not observable.

The contrapositive can be proven as follows. As defined in Definition~\ref{def:CriticalSet}, a critical set $\mathcal{C}^m$ consists of a maximal set of measurements which can be assigned to a line in $\mathcal{L}^m$ to connect two subgraphs of the power system graph, namely, $\mathcal{G}^m_1(\mathcal{N}^m_1,\mathcal{L}^m_1)$ and $\mathcal{G}^m_2(\mathcal{N}^m_2,\mathcal{L}^m_2)$, while preserving the ability to construct spanning trees over each of $\mathcal{G}^m_1(\mathcal{N}^m_1,\mathcal{L}^m_1)$ and $\mathcal{G}^m_2(\mathcal{N}^m_2,\mathcal{L}^m_2)$ using measurements in $\{\mathcal{M}^m_1\cup\mathcal{M}^m_2\}\setminus\mathcal{C}^m$. 
Consider that a number of measurements in the system are unavailable (i.e. attacked) so that the remaining measurements are not enough to include all critical sets in a maximum matching. Without loss of generality, we consider $\mathcal{C}^m$ to be an unmatched critical set.

When $\mathcal{C}^m$ is not part of a maximum matching over a CS-SMBG, then either 1) all of the measurements in $\mathcal{C}^m$ were removed (i.e. attacked and are not available), or 2) other measurements in the system were removed so that not enough measurements remain to match all critical sets (i.e. all measurements in $\mathcal{C}^m$ are matched to other critical sets). 

Now, if none of the measurements in $\mathcal{C}^m$ could be matched to $\mathcal{C}^m$ $\Rightarrow$ no measurement in $\mathcal{C}^m$ could be assigned to a line in $\mathcal{L}^m$ since a measurement can be assigned to one line at a time $\Rightarrow$ $\mathcal{G}^m_1(\mathcal{N}^m_1,\mathcal{L}^m_1)$ and $\mathcal{G}^m_2(\mathcal{N}^m_2,\mathcal{L}^m_2)$ could not be connected while preserving the observability within each of $\mathcal{G}^m_1(\mathcal{N}^m_1,\mathcal{L}^m_1)$ and $\mathcal{G}^m_2(\mathcal{N}^m_2,\mathcal{L}^m_2)$ (i.e., concurrently having a spanning tree over each of $\mathcal{G}^m_1$ and $\mathcal{G}^m_2$) $\Rightarrow$ a spanning tree over the power system graph $\mathcal{G}(\mathcal{N},\mathcal{L})$ could not be constructed through measurement assignments $\Rightarrow$ the power system is unobservable.
This then proves the contrapositive of this theorem and, hence, proves the theorem.   
\end{IEEEproof}

Theorem~\ref{theorem:ObservabilityGeneral} can be used to fully characterize observability attacks as follows. An observability attack is one in which measurements are removed (i.e. nodes from the right-side of a CS-SMBG) such that any maximum matching over this CS-SMBG would not include at least one critical set (i.e. nodes on the left-side of a CS-SMBG), which renders the system unobservable. This applies to any CS-SMBG, that is built from any set of critical sets (containing a critical set for each assigned measurement), that are derived from any original spanning tree $\mathcal{T}$. 
This, as a result, provides a general analytical characterization of observability attacks and enables prediction of the effect of the removal of a subset of measurements on the observability of the system. This enables identifying security indices such as the observability attack of lowest cardinality or the minimal set of measurements to remove in addition to a certain measurement to make the system unobservable.
In what follows, we focus on stealthy data injection attacks -- proving that they are a subset of observability attacks -- and we show how our provided analytical characterization of observability attacks enables analysis of various widely-studied stealthy data injection attack problems. To this end, we introduce sets of measurements, dubbed \emph{observability sets}, as follows, which are valuable for the analysis of data injection attacks which ensues.

\begin{definition}
For a CS-SMBG, an \emph{observability set} $\mathcal{S}\in\mathcal{M}$ is a set of measurements such that removing $\mathcal{S}$ leads a maximum matching over this CS-SMBG not to include a certain critical set.
\end{definition} 

In this regard, adding any measurement $s\in\mathcal{S}$, which was removed, back to the right-side of this bipartite graph will result in re-including the previously unmatched critical set in this maximum matching. In addition, removing all of $\mathcal{S}$ except for one measurement $s\in\mathcal{S}$ will not lead to excluding any critical set from a maximum matching over the CS-SMBG (resulting in a perfect matching). We note that multiple observability sets may exist for a CS-SMBG, as the removal of a different set of measurements may lead to the exclusion of a different critical set from a maximum matching over this graph. Here, we note that an observability set is associated with the CS-SMBG from which it was derived. A different CS-SMBG (resulting from a different set of critical sets, originating from a different original spanning tree) would result in different observability sets.

A \emph{union of observability sets} for a CS-SMBG is defined as a set of measurements composed of a number of observability sets such that, when each of these sets is successively removed, each such removal leads to excluding one additional critical set from being part of a maximum matching over the CS-SMBG. Adding back any of the removed measurements to the right-side of the bipartite graph will result in re-including one of the unmatched critical sets in the previously obtained maximum matching. 
Note that with the successive removal of observability sets, observability sets are defined based on the updated state of the CS-SMBG after the removal of a previous observability set at a previous step. In addition, at each step, multiple observability sets may exist. 
These observability sets play a crucial role in characterizing stealthy data injection attacks, as shown next. 

We next introduce stealthy data injection attacks and prove that they are a variant of our introduced observability attacks. This enables further studying and solving various problems related to SDIAs using our developed analytical tools. 
 
\section{Stealthy Data Injection Attacks}\label{sec:StealthyDIAs}

\subsection{Stealthy Data Injection Attacks}

Recalling the measurement-state equation in~(\ref{eq:MeasState}), data injection attacks aim at replacing the measurement vector, $\boldsymbol{z}$ by a manipulated measurement vector $\boldsymbol{z}^a=\boldsymbol{z}+\boldsymbol{a}$, where $\boldsymbol{a}\in\mathds{R}^{m}$ is the \emph{attack vector}, resulting in a new state estimate $\hat{\boldsymbol{x}}^a$. 
%
However, typically, the state estimation process is run in conjunction with what is known as a bad data detector and identifier (BDD). The BDD aims at detecting and identifying the presence of outliers in the collected data set, so that such outliers can be removed preventing them from affecting the estimation outcome. Such BDDs rely on the statistical analysis of what is known as measurement residuals, $\boldsymbol{r}$, defined as~\cite{Abur}:
\begin{align}\label{eq:Residuals}
\boldsymbol{\hat{z}}=\boldsymbol{H}\hat{\boldsymbol{x}}=\boldsymbol{S}\boldsymbol{z}, \ \boldsymbol{r}=\boldsymbol{z}-\hat{\boldsymbol{z}}=(\boldsymbol{I}_n-\boldsymbol{S})\boldsymbol{z}=\boldsymbol{W}\boldsymbol{z}, 
\end{align}
where $\boldsymbol{S}=\boldsymbol{H}(\boldsymbol{H}^T\boldsymbol{R}^{-1}\boldsymbol{H})^{-1}\boldsymbol{H}^T\boldsymbol{R}^{-1}$ and $\boldsymbol{W}=\boldsymbol{I}_n-\boldsymbol{S}$.

A statistical analysis on the residuals enables analysis of the magnitudes of the errors associated with each measurement, and hence, allows the identification of outliers~\cite{Abur}. Regarding data injection attacks, when data is added to certain measurements, the adversary aims at keeping the residuals unchanged, so that the attack cannot be detected by the BDD. Indeed, as shown in~\cite{liu1stdatainjection}, an attack vector that falls in the column-space of the Jacobian matrix $\boldsymbol{H}$, i.e. $\boldsymbol{a}=\boldsymbol{H}\boldsymbol{c}$, cannot be detected by residual statistical analysis. Indeed, for $\boldsymbol{a}=\boldsymbol{H}\boldsymbol{c}$,
\begin{flalign}\label{eq:ResidualwAtt}
\boldsymbol{r}^a&=\boldsymbol{W}(\boldsymbol{z}+\boldsymbol{a})=\boldsymbol{r}+\boldsymbol{W}\boldsymbol{a}
=\boldsymbol{r}+\boldsymbol{H}\boldsymbol{c}-\boldsymbol{H}\boldsymbol{c}=\boldsymbol{r}.&
\end{flalign}
  
As such, given the weighted least squares state estimation equation in~(\ref{eq:SE}), the attack vector $\boldsymbol{a}=\boldsymbol{H}\boldsymbol{c}$ generates an arbitrary new state estimate $\hat{\boldsymbol{x}}^a=\hat{\boldsymbol{x}}+\boldsymbol{c}$ by choosing the constant vector $\boldsymbol{c}$ without inducing any changes to the residual vector, as shown in~(\ref{eq:ResidualwAtt}). Such DIAs are, hence, stealthy and are referred to as stealthy DIAs. The ability of SDIAs to stealthily manipulate the state estimates poses various challenges to the operation of the grid. Hence, understanding and modeling such attacks is indispensable to the secure and sustainable operation of power systems. 
To this end, we next introduce a holistic graph-theoretic modeling of SDIAs that is based on the graph-theoretic modeling of observability attacks introduced in Section~\ref{sec:ObservabilityAtt}.     

\subsection{Graph-Theoretic Modeling of SDIAs}
The observability attacks and observability sets introduced in Section~\ref{sec:ObservabilityAtt} provide the basis for a graph-theoretic interpretation of SDIAs as will be shown in Theorem~\ref{Theorem:sDIAObservability}. However, before introducing and proving Theorem~\ref{Theorem:sDIAObservability}, we introduce a preliminary lemma which will be used in the proof of Theorem~\ref{Theorem:sDIAObservability}. 

\begin{lemma}\label{lemma:sDIARemovalifUnobservable}
If a DIA is stealthy (i.e. $\boldsymbol{a}=\boldsymbol{H}\boldsymbol{c}$), then removing the attacked measurements renders the system unobservable. 
\end{lemma}
\begin{IEEEproof}
Since the attack vector $\boldsymbol{a}$ is stealthy, then $\boldsymbol{a}=\boldsymbol{H}\boldsymbol{c}$. Since $\boldsymbol{H}$ is of full rank, then the only solution to $\boldsymbol{H}\boldsymbol{c}=0$ is $\boldsymbol{c}=\boldsymbol{0}$. Hence, $\boldsymbol{a}=\boldsymbol{H}\boldsymbol{c}$ has zero and nonzero elements for $\boldsymbol{c}\neq\boldsymbol{0}$. Now, if all of the rows of $\boldsymbol{H}$ corresponding to nonzero elements of $\boldsymbol{a}$ are removed to form matrix $\boldsymbol{H}_{\textrm{new}}$, then, this results in $\boldsymbol{a}_{\textrm{new}}=\boldsymbol{H}_{\textrm{new}}\boldsymbol{c}=\boldsymbol{0}$ for $\boldsymbol{c}\neq\boldsymbol{0}$. Hence, $\boldsymbol{H}_{\textrm{new}}$ is not of full rank and the power system whose Jacobian matrix is given by $\boldsymbol{H}_{\textrm{new}}$ is unobservable. Therefore, when the attack is stealthy, removing the attacked measurements renders the system unobservable.    
\end{IEEEproof}

Here we note, that the result of Lemma~\ref{lemma:sDIARemovalifUnobservable}, provides a one directional relation stating that if $\boldsymbol{a}=\boldsymbol{H}\boldsymbol{c}$, i.e. the attack is stealthy, then the removal of the nonzero elements of $\boldsymbol{a}$, i.e. the attacked measurements, causes the system to be unobservable. However, the reverse direction does not always hold true. Indeed, the reverse statement of Lemma~\ref{lemma:sDIARemovalifUnobservable} states that, if removing a set of measurements renders the system unobservable, then this guarantees that a stealthy DIA can be constructed which targets all of these measurements, and only these measurements. We next provide a counter example which proves that this reverse statement does not hold true. In this regard, we consider the Jacobian matrix $\boldsymbol{H}$ to be represented as follows: $\boldsymbol{H}=\left[ {\begin{array}{c} \boldsymbol{H}_0 \\\boldsymbol{H}_1 \end{array}} \right]$. We let $\mathcal{M}_0$ and $\mathcal{M}_1$ represent the subset of measurements corresponding to the rows of $\boldsymbol{H}_0$ and $\boldsymbol{H}_1$, respectively. Consider $\mathcal{M}_0$ to contain one critical measurement, i.e., one row of $\boldsymbol{H}_0$ is independent of all of the other rows of $\boldsymbol{H}$. As such, removing the subset of measurements $\mathcal{M}_0$ renders the system unobservable. In addition, consider two measurements $m_0\in\mathcal{M}_0$ and $m_1\in\mathcal{M}_1$ such as $m_0$ measures the power flow from bus $i$ to bus $j$ and $m_1$ measures the power flow from bus $j$ to bus $i$ (i.e., $m_0$ and $m_1$ are installed on the same transmission line but measure the flow in two opposite directions). In this regard, let $\boldsymbol{h}_0$ and $\boldsymbol{h}_1$ correspond to the rows of $m_0$ and $m_1$ in, respectively, $\boldsymbol{H}_0$ and $\boldsymbol{H}_1$. Then, we have\footnote{Since $P_{ij}=-P_{ji}$, where $P_{ij}$ and $P_{ji}$ are the real power flow from bus $i$ to bus $j$ and from bus $j$ to bus $i$, respectively, over the same transmission line.} $\boldsymbol{h}_0=-\boldsymbol{h}_1$. 
As a result, one cannot find a stealthy attack vector $\boldsymbol{a}=\left[ {\begin{array}{c} \boldsymbol{a}_0 \\\boldsymbol{a}_1 \end{array}} \right]=\left[ {\begin{array}{c} \boldsymbol{H}_0 \\\boldsymbol{H}_1 \end{array}} \right]\boldsymbol{c}$, in which all the elements of $\boldsymbol{a}_0$ are nonzero and all the elements of $\boldsymbol{a}_1$ are zero, since if $\boldsymbol{h}_0\boldsymbol{c}\neq0$, then $\boldsymbol{h}_1\boldsymbol{c}\neq0$, due to the fact that $\boldsymbol{h}_0=-\boldsymbol{h}_1$. This implies that for the attack to target all the measurements in $\mathcal{M}_0$ and be stealthy, this attack must also target measurements in $\mathcal{M}_1$. Otherwise, this attack must be limited to a strict subset of $\mathcal{M}_0$ and may not target all the measurements in $\mathcal{M}_0$. As a result, even though removing the measurements in $\mathcal{M}_0$ renders the system unobservable, one cannot necessarily construct a stealthy attack vector that only targets all the measurements in $\mathcal{M}_0$. Hence, this provides a counter example of the reverse statement of Lemma~\ref{lemma:sDIARemovalifUnobservable} proving that this reverse statement does not always hold true.

\begin{theorem}\label{Theorem:sDIAObservability}
A DIA is stealthy if and only if the attacked measurements constitute a union of observability sets over a certain CS-SMBG.
\end{theorem}
\begin{IEEEproof}
We begin by proving that when the attacked measurements (i.e. nonzero elements of the attack vector $\boldsymbol{a}$) constitute a union of observability sets, then $\boldsymbol{a}$ is stealthy (i.e. $\boldsymbol{a}$ can be represented as $\boldsymbol{a}=\boldsymbol{H}\boldsymbol{c}$). 
As shown in Theorem~\ref{theorem:ObservabilityGeneral}, when an observability set (equivalently, a union of observability sets) is removed, the system is unobservable. Hence, consider an observability set $\mathcal{S}$ which has been removed. Let $\boldsymbol{H}^{(-\mathcal{S})}$ be the system's Jacobian matrix without the measurements in $\mathcal{S}$ and let $\mathcal{C}$ be the critical set which cannot be part of a maximum matching over the CS-SMBG when $\mathcal{S}$ is removed. Since the system is unobservable when removing $\mathcal{S}$, 
$\boldsymbol{H}^{(-\mathcal{S})}\boldsymbol{y}=0$ for a $\boldsymbol{y}\neq\boldsymbol{0}$. However, the addition of any measurement $s\in\mathcal{S}$ will reinclude $\mathcal{C}$ in the maximum matching over the CS-SMBG, and hence, reconnect the tree. As such, let $\boldsymbol{H}^{(-\mathcal{S})+(k)}$ correspond to $\boldsymbol{H}^{(-\mathcal{S})}$ with the addition of a row corresponding to a measurement $k\in\mathcal{S}$. In this regard, since the system is rendered observable, $\boldsymbol{H}^{(-\mathcal{S})+(k)}$ is of full rank and $\boldsymbol{H}^{(-\mathcal{S})+(k)}\boldsymbol{y}$ will have one nonzero element corresponding to the row of $\boldsymbol{H}^{(-\mathcal{S})+(k)}$ pertaining to the added measurement $k$. This procedure can be repeated for all $k\in\mathcal{S}$. As such, adding the rows corresponding to $\mathcal{S}$ back to the Jacobian matrix results in $\boldsymbol{b}=\boldsymbol{H}\boldsymbol{y}$ in which only the elements of $\boldsymbol{b}$ corresponding to measurements in $\mathcal{S}$ are nonzero. As a  result, $\boldsymbol{a}=\boldsymbol{b}$ is an attack vector in which only the observability set $\mathcal{S}$ is attacked and is proven to be stealthy.

Now, we prove that, when an attack is stealthy, i.e. $\boldsymbol{a}=\boldsymbol{H}\boldsymbol{c}$, then the nonzero elements of $\boldsymbol{a}$ correspond to a union of observability sets. 
In this regard, from Lemma~\ref{lemma:sDIARemovalifUnobservable}, we know that removing the nonzero elements of $\boldsymbol{a}=\boldsymbol{H}\boldsymbol{c}$ will render the system unobservable, which implies that the nonzero elements of $\boldsymbol{a}$ contain at least one observability set. Let $\mathcal{S}$ denote this observability set, and let $\boldsymbol{H}^{(-\mathcal{S})}$ be the system's Jacobian matrix without the measurements in $\mathcal{S}$. Removing $\mathcal{S}$ will lead to two subsystems each of which is fully observable (i.e. it will split the spanning tree, $\mathcal{T}$, into two subtrees each of which spans its own subgraph). Let $\boldsymbol{H}_1$ and $\boldsymbol{H}_2$ be the Jacobian matrices of each of these two subsystems (we denote these subsystems as subsystem 1 and subsystem 2) and let $\boldsymbol{a}_1$ and $\boldsymbol{a}_2$ correspond to the portions of $\boldsymbol{a}$ (excluding the measurements of the previously removed observability set) corresponding to the measurements in $\boldsymbol{H}_1$ and $\boldsymbol{H}_2$, respectively. In addition, let $\boldsymbol{c}_1$ and $\boldsymbol{c}_2$ correspond to the portions of $\boldsymbol{c}$ pertaining to nodes in subsystem 1 and subsystem 2, respectively. Now, if $\boldsymbol{a}_i$ for $i\in\{1,2\}$ has nonzero elements, this implies that removing these elements will make subsystem $i$ unobservable, which implies that the nonzero elements of $\boldsymbol{a}_i$ contain an observability set. 
Following this same logic, removing this observability set will subsequently split subsystem $i$ into two subsystems, each of which is observable. This process can be continued recursively until no measurement $m$ corresponding to a nonzero element of $\boldsymbol{a}$ remains. Hence, this shows that when $\boldsymbol{a}=\boldsymbol{H}\boldsymbol{c}$, then the nonzero elements of $\boldsymbol{a}$ correspond to a union of observability sets.

This proves both directions of the theorem, and hence, concludes the proof.  
\end{IEEEproof}

Theorem~\ref{Theorem:sDIAObservability} provides an analytical graph-theoretic modeling of SDIAs using the fundamentals of observability attacks introduced in Section~\ref{sec:ObservabilityAtt}. This enables a fundamental understanding of SDIAs since it allows the characterization of the subset of measurements which would be compromised as part of an SDIA and hence enables defense against such attacks. In addition, this analytical characterization of SDIAs enables a more in-depth analysis of such integrity attacks and allows a characterization of analytical solutions to a wide-range of well-studied problems in this field, as will be explored in Section~\ref{subsec:UnifiedSolutionsSDIAs}.  

\begin{figure}[t!]
  \begin{center}
   \vspace{-0.35cm}
    \includegraphics[width=8.5cm]{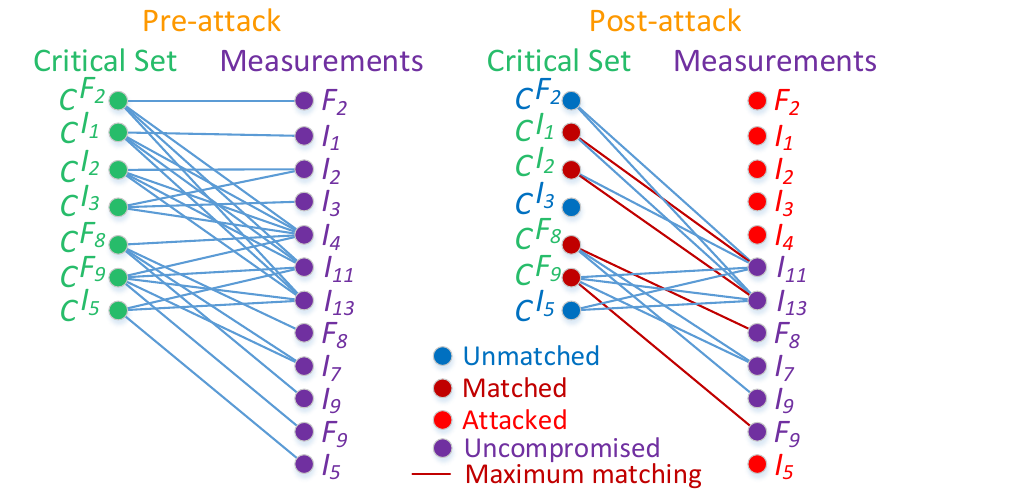}
    \vspace{-0.45cm}
    \caption{\label{fig:CriticalSetSystmMeasBipartitesDIA} Critical sets - system measurements maximum matching and SDIAs.}
  \end{center}\vspace{-0.65cm}
\end{figure}

\begin{example}
As an illustrative example of the result\footnote{In this example, we index the measurements in Fig.~\ref{fig:IEEE14Bus} from $1$ to $17$ in an incremental manner based on the following order $(F_2,\, F_8,\, F_9,\, F_{15},\, I_1,\, I_2,\, I_3,\, I_4,\, I_5,\, I_6,\, I_7,\, I_9,\, I_{11},\, I_{12},\, I_{13},\, F_{17},\, F_{19})$.} in Theorem~\ref{Theorem:sDIAObservability}, we consider the IEEE 14-bus system, shown in Fig.~\ref{fig:IEEE14Bus}, whose line transmission data can be found in~\cite{MATPOWER}. We consider the stealthy attack $\boldsymbol{a}=\boldsymbol{H}\boldsymbol{c}$ with $\boldsymbol{c}=[1,0,...,0]^T$, which corresponds to having the attack vector equal to the first column of the Jacobian matrix $\boldsymbol{H}$ given by $\boldsymbol{H}(:,1)=[-16.9,0,0,0,-16.9,33.37,-5.05, -5.67,-5.75,\textrm{zeros}(1,8)]^T$. This attack consists of attacking measurement indices $\{1, 5, 6, 7, 8, 9\}$ which correspond to $\{F_2, I_1, I_2, I_3, I_4, I_5\}$. In this respect, we next verify whether this attack is stealthy, following Theorem~\ref{Theorem:sDIAObservability}. 
To this end, Fig.~\ref{fig:CriticalSetSystmMeasBipartitesDIA} shows a portion of the CS-SMBG that is relevant to the attacked measurements. The post-attack portion of Fig.~\ref{fig:CriticalSetSystmMeasBipartitesDIA} marks the nodes corresponding to measurements $\{F_2, I_1, I_2, I_3, I_4, I_5\}$, on the right-side of the bipartite graph, as attacked (following the attack vector $\boldsymbol{a}$). As a result, all the edges connecting these nodes to the critical sets on the left-side of the bipartite graph are removed. Then, building a maximum matching over the post-attack bipartite graph shows that, indeed, not all the critical sets are matched. Hence, the removed measurements lead to a maximum matching that does not include all critical sets. Furthermore, the addition of a node corresponding to any of the attacked measurements, i.e. $\{F_2, I_1, I_2, I_3, I_4, I_5\}$, would lead to reincluding one of the unmatched critical sets $\{\mathcal{C}^{F_2}, \mathcal{C}^{I_3}, \mathcal{C}^{I_5}\}$ in the maximum matching. This implies that the attack consists of a union of observability sets which implies that the attack is stealthy. 
\end{example}                      

%

\subsection{Unified Analysis of Diverse SDIA Problems}\label{subsec:UnifiedSolutionsSDIAs} 
Theorem~\ref{Theorem:sDIAObservability} provides a unified basis for studying various SDIA problems from a graph-theoretic perspective, as show next. 

SDIA analyses can be categorized based on whether the focus is on modeling the attack or the defense strategies. As such, we first present two problems focusing on \emph{modeling attack strategies} followed by two problems focusing on the derivation of \emph{defense strategies} to thwart SDIAs.

\subsubsection{Modeling SDIA Attack Strategies}

Modeling SDIA attack strategies enables a vulnerability assessment of the system and allows anticipating sophisticated attack strategies which can target the system. This, in turn, allows the derivation of adequate defense strategies to thwart such attacks. 
We next focus on two problems which aim at modeling potential attack strategies.   


\emph{Problem 1:} If measurement $k\in\mathcal{M}$ is attacked, what is a minimal set of measurements which must be attacked along with $k$ for the attack to be stealthy? In other words, \emph{Problem 1} seeks a solution to the following optimization problem:
\begin{align}
\underset{\boldsymbol{c}}\min ||\boldsymbol{H}\boldsymbol{c}||_0,\nonumber\\
\textrm{subject to: } \boldsymbol{H}(k,:)\boldsymbol{c}=1.
\end{align} 

\emph{Problem 1} has been proposed in~\cite{DataInjSecIndex} and studied in~\cite{SparsestAttackCDC} and~\cite{DIAsSecIndexTAC}. The derived solution in~\cite{SparsestAttackCDC} is based on an approximate relaxation method while the solution in~\cite{DIAsSecIndexTAC} focuses on the special case assuming that the measurement set consists of all injection measurements at all buses and all line flow measurements at all transmission lines. Instead, here, we provide a general graph-theoretic characterization of a solution to this problem using our developed framework. We note that the goal of this analysis here is not to determine and analyze the computational efficiency of deriving a solution to \emph{Problem 1} but rather to analytically characterize a solution to this problem using our developed graph-theoretic framework\footnote{Exact complexity analysis can be an interesting subject for future work.}.


An analytical graph-theoretic solution to \emph{Problem 1} is characterized in Theorem~\ref{prop:KStealthyAttack}.

\begin{theorem}\label{prop:KStealthyAttack}
A stealthy attack of smallest cardinality containing measurement $k$ corresponds to attacking the measurements of the critical set of lowest cardinality which contains $k$.
\end{theorem}
\begin{IEEEproof}
First, we show that the attack containing the critical set of lowest cardinality containing $k$ is, indeed, stealthy. Then, we prove that this attack is a stealthy attack containing $k$ that has a minimum cardinality. 

By Theorem~\ref{Theorem:sDIAObservability}, for the attack to be stealthy, the removal of the attacked measurements must lead 
a maximum matching over a CS-SMBG not to include all the critical sets (i.e. all the left-side nodes of the bipartite graph). In other words, the attack must be composed of a union of obseravbility sets. In this respect, removing the critical set containing $k$ that is of smallest cardinality is, indeed, stealthy since removing an entire critical set will disconnect the node corresponding to this critical set (on the left-hand side of the CS-SMBG) from the right-side of the bipartite graph, which prevents this critical measurement from being part of any maximum matching. 

Next, we prove that there are no stealthy attacks containing $k$ that have a smaller cardinality. In this regard, for an attack containing $k$ to be stealthy, it must prevent a critical set, in which $k$ exists, from being part of 
a maximum matching over a certain CS-SMBG. A critical set would be excluded from 
such a maximum matching in two cases: 1) if all the measurements in this critical set are attacked, or 
2) if all the measurements in this critical set are matched to other critical sets.

In the first case, considering attacking all the measurements in a critical set, then attacking the critical set that has the fewest number of measurements -- as stated in this theorem -- corresponds to the minimum cardinality attack for this specific CS-SMBG. 
As for the second case, if a measurement $k'$ in a critical set containing $k$ is matched -- as part of a maximum matching -- to another critical set (we denote this set by $\mathcal{C}^p$), then measurement $p$ must be attacked since, otherwise, $\mathcal{C}^p$ would have been matched to $p$ sparing $k$ to be matched to another critical set to maximize the cardinality of the matching. In other words, matching a critical set $\mathcal{C}^p$ with a measurement $k'\ne p$ while $p$ is not attacked is contradictory to the assumption that this matching is maximum. As a result, for a critical set $\mathcal{C}$, such that $k\in\mathcal{C}$, to be discarded from a maximum matching, every measurement in $\mathcal{C}$ must be matched to another critical set. This implies that at least one measurement of each of these critical sets is attacked. Thus, the number of attacked measurements will be at least equal to the number of measurements within $\mathcal{C}$ for the attack to be stealthy. Consequently, for a certain CS-SMBG, attacking the critical set that has the fewest number of measurements and which contains $k$ is a SDIA containing $k$ having the lowest cardinality. 
As a result, considering all possible CS-SMBGs, obtained from any set of critical sets that were derived from any original tree, and considering the minimum-cardinality critical set containing $k$ in each one of these CS-SMBGs leads to the following result: a stealthy attack containing $k$ that has the lowest cardinality consists of attacking only the measurements of the critical set containing $k$ that has the lowest cardinality, considering all such critical sets derived from any possible original spanning tree.     
\end{IEEEproof}

Here, one can also conclude that the solution to \emph{Problem 1} may not be unique since a critical set containing $k$ of minimum cardinality may not be unique. Thus, any critical set containing $k$, whose cardinality is less than or equal to any other critical set containing $k$, is a solution to \emph{Problem 1}. 
In addition, a numerical derivation to the solution of \emph{Problem 1} requires exploring all possible spanning trees over $\mathcal{G}$. This could be performed, for example, based on a repeated application of the spanning tree building algorithms proposed in~\cite{TopologicalObs1,ObservabulityAlgorithmGTMatroid,ObservabilityGTAlgorithm}, among other techniques. However, our goal here is to provide a graph-theoretic characterization to the solution of \emph{Problem 1}, rather than proposing algorithms to numerically obtain this solution. This latter end goal would highly complement the results of the current work and is left to be explored in future extensions.

Characterizing the solution to \emph{Problem 1} will also facilitate solving another key SDIA problem, referred to as \emph{Problem 2}, and stated as follows. 

\emph{Problem 2:} What is the SDIA with the lowest cardinality? In other words, which SDIA is a solution to:
\begin{align}
\underset{\boldsymbol{a}}\min ||\boldsymbol{a}||_0,\nonumber\\
\textrm{subject to: } \boldsymbol{a}=\boldsymbol{H}\boldsymbol{c}.
\end{align}  


A solution to \emph{Problem 2} is provided in Proposition~\ref{prop:SmallestStealthyAttack}.

\begin{prop}\label{prop:SmallestStealthyAttack}
A stealthy attack of lowest cardinality consists of attacking the smallest critical set.
\end{prop}
\begin{IEEEproof}
This proof follows directly from the proof of Theorem~\ref{prop:KStealthyAttack}. Indeed, since the stealthy attack containing measurement $k$ that is of smallest cardinality corresponds to the critical set of lowest cardinality containing $k$, searching for a global stealthy attack of lowest cardinality can be limited to only critical sets, considering the set of critical sets originating from each possible original spanning tree. Based on this fact, a stealthy attack of lowest cardinality is the one in which the measurements in the critical set of lowest cardinality are the only measurements that are attacked (the only measurements having nonzero corresponding elements in the attack vector $\boldsymbol{a}$). 
\end{IEEEproof}

As multiple critical sets may have the same cardinality, the solution to \emph{Problem 2} may not be unique.

\subsubsection{Modeling SDIA Defense Strategies} Using our introduced graph-theoretic framework, two fundamental widely-studied problems for defending the system against SDIAs are presented and investigated, next, in \emph{Problem 3} and \emph{Problem 4}.

\emph{Problem 3:} What is the minimum set of measurements that must be protected (i.e. made immune to SDIAs) to guarantee no SDIAs can be successful?


The solution to \emph{Problem 3} is presented in Theorem~\ref{prop:PerfectDefense}.

\begin{theorem}\label{prop:PerfectDefense}
A minimum set of measurements that must be protected to guarantee that no SDIA can be successful consists of protecting all measurements in $\mathcal{M}^A$, i.e. all measurements that are part of the original assignment function forming a spanning tree over the power system. 
\end{theorem}

\begin{IEEEproof}
%
Protecting all the measurements in $\mathcal{M}^A$ will guarantee that these measurements will be part of the Jacobian matrix $\boldsymbol{H}$. Since these measurements form a spanning tree over the power system, their rows in $\boldsymbol{H}$ are linearly independent. As such, let $\boldsymbol{H}^A$ be the Jacobian matrix corresponding only to measurements in $\mathcal{M}^A$, then $\boldsymbol{H}^A\boldsymbol{c}=\boldsymbol{0}$ has no solution other than $\boldsymbol{c}=\boldsymbol{0}$. The rows of $\boldsymbol{H}^A$ are a subset of the rows of $\boldsymbol{H}$. As such, one cannot find an attack vector $\boldsymbol{a}=\boldsymbol{H}\boldsymbol{c}$ such that all the elements of $\boldsymbol{a}$ corresponding to the rows of $\boldsymbol{H}^A$ are zero. Hence, one cannot find a stealthy attack $\boldsymbol{a}=\boldsymbol{H}\boldsymbol{c}$ which does not attack the measurements in $\mathcal{M}^A$. As a result, protecting these measurements will guarantee that no stealthy attack can be carried out. This set is a minimum set since if one measurement $m\in\mathcal{M}^A$ is not protected, a SDIA can be successfully launched by attacking the critical set $\mathcal{C}^m$ (by definition of a critical set, no measurement in $\mathcal{C}^m$ is part of $\mathcal{M}^A$ except for $m$, which makes attacking $\mathcal{C}^m$ valid). Hence, protecting the system against any SDIA requires at least defending $N-1$ linearly independent rows of $\boldsymbol{H}$, which constitute the assigned measurements $\mathcal{M}^A$ of $\mathcal{T}$.   
\end{IEEEproof}

Note that, the solution to \emph{Problem 3} may not be unique, since the subset of measurements which can be assigned to lines to form a spanning tree over $\mathcal{G}(\mathcal{N},\mathcal{L})$ is not necessarily unique. However, by Theorem~\ref{prop:PerfectDefense}, we can find one of the solutions to \emph{Problem 3}. Starting from a different spanning tree, and following the same steps, would result in another valid solution to \emph{Problem 3}. All these solutions have the same minimum cardinality.

Characterizing a solution to Problem 3 provides important information regarding the size of investments needed to make a power system immune to SDIAs. In this regard, regardless of how high the number of measurements in an $N$-bus system is, the number of measurements that must be protected to render the system immune to SDIAs is always equal to $N-1$. 

\begin{example}
Applying the results in Theorem~\ref{prop:PerfectDefense} to our treated IEEE 14-bus system case analysis, protecting the measurements in the first column of Table~\ref{Tab:CriticalSets} is a set of measurements of minimum cardinality which when protected renders the IEEE 14-bus system in Fig.~\ref{fig:IEEE14Bus} immune to any SDIA.
\end{example}   

Theorem~\ref{Theorem:sDIAObservability} can be used to characterize a solution to \emph{Problem 4} which was proposed in~\cite{Poor} and which is presented next. The work in~\cite{Poor} focused on deriving an $l_1$ relaxation and an approximate numerical solution of the corresponding optimization problem. Here, we focus on an analytical analysis of this problem.

\emph{Problem 4:} What is a minimum set of measurements to protect as to force the attacker to manipulate at least $\tau_a$ measurements to stay stealthy?


A solution to \emph{Problem 4} is presented in Proposition~\ref{prop:DefenseThresholdTau}.

\begin{prop}\label{prop:DefenseThresholdTau}
Consider $\mathcal{M}_{\tau_a}^{\textrm{CS-SMBG}}$ to be a set of measurements including one distinct measurement from each critical set whose cardinality is lower than $\tau_a$ for a certain CS-SMBG. Then, a minimum set of measurements to protect so that no attack with cardinality $||\boldsymbol{a}||_0<\tau_a$ can be stealthy, corresponds to finding a minimum set of measurements whose elements include, one $\mathcal{M}_{\tau_a}^{\textrm{CS-SMBG}}$ for each possible CS-SMBG.
\end{prop}

\begin{IEEEproof}
For a certain CS-SMBG and following from Theorem~\ref{Theorem:sDIAObservability}, for an attack with cardinality lower than $\tau_a$ to be stealthy, the attacked measurements must constitute a union of observability sets, whose cardinality must be lower than $\tau_a$. Hence, if all critical sets whose cardinalities are lower than $\tau_a$ are guaranteed to be matched in any maximum matching over this CS-SMBG, attacking a union of observability sets of cardinality lower than $\tau_a$ would be made impossible. 
This could only be achieved by securing one distinct measurement in each of these sets. 
Thus, in a certain CS-SMBG, protecting one distinct measurement from each critical set whose cardinality is less than $\tau_a$ is a necessary condition for thwarting all possible SDIAs with cardinality lower than $\tau_a$, which could target this CS-SMBG. 
Now, considering all possible spanning trees and all the possible sets of critical sets which could be generated from each original spanning tree (giving rise to different CS-SMBGs), preventing all possible SDIAs whose 
$||\boldsymbol{a}||_0<\tau_a$ requires protecting one distinct measurement from each critical set whose cardinality is less than $\tau_a$ within each CS-SMBG. 
Hence, for each CS-SMBG, a distinct measurement must be protected from each critical set whose cardinality is lower than $\tau_a$. 
For a CS-SMBG, let $\mathcal{M}_{\tau_a}^{\textrm{CS-SMBG}}$ be a set of measurements containing one distinct measurement from each critical set whose cardinality is lower than $\tau_a$. We note that different $\mathcal{M}_{\tau_a}^{\textrm{CS-SMBG}}$ can be obtained for each CS-SMBG. As such, considering all possible CS-SMBG, and all possible $\mathcal{M}_{\tau_a}^{\textrm{CS-SMBG}}$ for each CS-SMBG, the solution to \emph{Problem 4} consists of finding a minimum set of measurements whose elements include one $\mathcal{M}_{\tau_a}^{\textrm{CS-SMBG}}$ for each possible CS-SMBG. 

\end{IEEEproof}

Considering the set of measurements $\mathcal{V}^{\textrm{CS-SMBG}}$ to be composed of the different $\mathcal{M}_{\tau_a}^{\textrm{CS-SMBG}}$ for each possible CS-SMBG, and considering $\mathcal{U}$ to be the union of all such $\mathcal{V}^{\textrm{CS-SMBG}}$, solving \emph{Problem 4} corresponds to the known \emph{``hitting set problem''}~\cite{HittingSetProbPaper,Karp1972}, in which the universe is given by $\mathcal{U}$, the subsets are given by $\mathcal{V}^{\textrm{CS-SMBG}}$, and the elements within each $\mathcal{V}^{\textrm{CS-SMBG}}$ are the collection of the possible $\mathcal{M}_{\tau_a}^{\textrm{CS-SMBG}}$. This problem is equivalent to the \emph{``vertex cover problem''} and is a known NP-hard problem~\cite{HittingSetProbPaper,Karp1972}. Approximate methods for the ``hitting set problem'' are discussed in~\cite{HittingSetProbPaper} and the references therein. 



Hence, using the proposed graph-theoretic framework enables us to characterize the solutions to various well-studied SDIA problems. 


%

\section{Conclusion and Future Outlook}\label{sec:Conclusion}
In this paper, we have introduced a novel graph-theoretic framework which enables a fundamental modeling of observability attacks targeting power systems and have proven that the widely-studied stealthy data injection attacks are a special case of such observability attacks. Based on this proposed framework, we have characterized analytical solutions to various central observability and data injection attack problems focusing on the sparsest SDIA, the sparsest SDIA including a certain measurement, the minimum set of measurements to defend to thwart all possible SDIAs, and the minimum set of measurements whose defense guarantees that no DIA below a certain cardinality can be stealthy. 

The proposed graph-theoretic framework provides a general analytical tool using which a wide set of key observability attacks and data injection attacks problems can be modeled and analyzed, and is not limited to the set of problem examples which are studied in this paper. 
In addition, the ability to analytically characterize attack and defense policies using the proposed framework allows studying problems that involve interactions between attackers and defenders from a game-theoretic perspective. Such analyses can account for the opponent's potential attack or defense strategies when designing, respectively, defense policies or attack vectors. As a result, such analyses allow the modeling and investigation of practical competitive attack vs. defense settings. This enables studying the effects of sophisticated observability attacks and data injection attacks on the system as well as the impact of proposed defense strategies within various application domains such as electricity markets, congestion management, and contingency analysis, among others, thus taking the application of our framework beyond the domain of power systems which motivated this study.         

\def\baselinestretch{0.78}
\bibliographystyle{IEEEtran}
\bibliography{reference}

\begin{thebibliography}{10}
\providecommand{\url}[1]{#1}
\csname url@samestyle\endcsname
\providecommand{\newblock}{\relax}
\providecommand{\bibinfo}[2]{#2}
\providecommand{\BIBentrySTDinterwordspacing}{\spaceskip=0pt\relax}
\providecommand{\BIBentryALTinterwordstretchfactor}{4}
\providecommand{\BIBentryALTinterwordspacing}{\spaceskip=\fontdimen2\font plus
\BIBentryALTinterwordstretchfactor\fontdimen3\font minus
  \fontdimen4\font\relax}
\providecommand{\BIBforeignlanguage}[2]{{%
\expandafter\ifx\csname l@#1\endcsname\relax
\typeout{** WARNING: IEEEtran.bst: No hyphenation pattern has been}%
\typeout{** loaded for the language `#1'. Using the pattern for}%
\typeout{** the default language instead.}%
\else
\language=\csname l@#1\endcsname
\fi
#2}}
\providecommand{\BIBdecl}{\relax}
\BIBdecl

\bibitem{SGSecSurvey1}
Y.~Mo, T.~H.~J. Kim, K.~Brancik, D.~Dickinson, H.~Lee, A.~Perrig, and
  B.~Sinopoli, ``Cyber--physical security of a smart grid infrastructure,''
  \emph{Proceedings of the IEEE}, vol. 100, no.~1, pp. 195--209, Jan 2012.

\bibitem{liu1stdatainjectionConf}
Y.~Liu, P.~Ning, and M.~K. Reiter, ``False data injection attacks against state
  estimation in electric power grids,'' in \emph{Proc. 16th ACM Conference on
  Computer and Communications Security}, Chicago, Illinois, USA, November 2009,
  pp. 21--32.

\bibitem{SanjabSaadJournal}
A.~Sanjab and W.~Saad, ``Data injection attacks on smart grids with multiple
  adversaries: A game-theoretic perspective,'' \emph{IEEE Transactions on Smart
  Grid}, vol.~7, no.~4, pp. 2038--2049, July 2016.

\bibitem{SanjabSaad}
------, ``Smart grid data injection attacks: To defend or not?'' in \emph{Proc.
  IEEE International Conference on Smart Grid Communications (SmartGridComm)},
  Nov 2015, pp. 380--385.

\bibitem{LeXie}
L.~Xie, Y.~Mo, and B.~Sinopoli, ``Integrity data attacks in power market
  operations,'' \emph{IEEE Transactions on Smart Grid}, vol.~2, no.~4, pp.
  659--666, Dec 2011.

\bibitem{AttDetDorfler}
F.~Pasqualetti, F.~Dorfler, and F.~Bullo, ``Attack detection and identification
  in cyber-physical systems,'' \emph{IEEE Transactions on Automatic Control},
  vol.~58, no.~11, pp. 2715--2729, Nov 2013.

\bibitem{RobustResilientCPSPwrSysBasar}
Q.~Zhu and {T. Ba\c{s}ar}, ``Robust and resilient control design for
  cyber-physical systems with an application to power systems,'' in \emph{Proc.
  50th IEEE Conference on Decision and Control and European Control Conference
  (CDC-ECC)}, Dec 2011, pp. 4066--4071.

\bibitem{Abur}
A.~Abur and A.~G. Exposito, \emph{Power System State Estimation: Theory and
  Implementation}.\hskip 1em plus 0.5em minus 0.4em\relax New York: Marcel
  Dekker, 2004.

\bibitem{woodwollenberg}
A.~J. Wood and B.~F. Wollenberg, \emph{Power Generation, Operation, and
  Control}.\hskip 1em plus 0.5em minus 0.4em\relax John Wiley \& Sons, 2012.

\bibitem{CritSet3}
J.~B.~A. London, L.~F.~C. Alberto, and N.~G. Bretas, ``Network observability:
  identification of the measurements redundancy level,'' in \emph{Proc.
  International Conference on Power System Technology (PowerCon)}, vol.~2,
  2000, pp. 577--582.

\bibitem{DataInjGTKTuple}
K.~C. Sou, H.~Sandberg, and K.~H. Johansson, ``Computing critical $k$-tuples in
  power networks,'' \emph{IEEE Transactions on Power Systems}, vol.~27, no.~3,
  pp. 1511--1520, Aug 2012.

\bibitem{DataInjGTKosut}
O.~Kosut, L.~Jia, R.~J. Thomas, and L.~Tong, ``Malicious data attacks on the
  smart grid,'' \emph{IEEE Transactions on Smart Grid}, vol.~2, no.~4, pp.
  645--658, Dec 2011.

\bibitem{DataInjSecIndex}
H.~Sandberg, A.~Teixeira, and K.~H. Johansson, ``On security indices for state
  estimators in power networks,'' in \emph{First Workshop on Secure Control
  Systems (SCS)}, 2010.

\bibitem{SparsestAttackCDC}
K.~C. Sou, H.~Sandberg, and K.~H. Johansson, ``Electric power network security
  analysis via minimum cut relaxation,'' in \emph{Proc. 50th IEEE Conference on
  Decision and Control and European Control Conference}, Dec 2011, pp.
  4054--4059.

\bibitem{DIAsSecIndexTAC}
J.~M. Hendrickx, K.~H. Johansson, R.~M. Jungers, H.~Sandberg, and K.~C. Sou,
  ``Efficient computations of a security index for false data attacks in power
  networks,'' \emph{IEEE Transactions on Automatic Control}, vol.~59, no.~12,
  pp. 3194--3208, Dec 2014.

\bibitem{Poor}
T.~Kim and H.~V. Poor, ``Strategic protection against data injection attacks on
  power grids,'' \emph{IEEE Transactions on Smart Grid}, vol.~2, no.~2, pp.
  326--333, June 2011.

\bibitem{TajerPoor}
S.~Cui, Z.~Han, S.~Kar, T.~T. Kim, H.~V. Poor, and A.~Tajer, ``Coordinated
  data-injection attack and detection in the smart grid: A detailed look at
  enriching detection solutions,'' \emph{IEEE Signal Processing Magazine},
  vol.~29, no.~5, pp. 106--115, Sept 2012.

\bibitem{CritSet3Journal}
J.~London, A.~Bretas, and N.~Bretas, ``Algorithms to solve qualitative problems
  in power system state estimation,'' \emph{International Journal of Electrical
  Power and Energy Systems}, vol.~26, no.~8, pp. 583 -- 592, 2004.

\bibitem{MminusKRobustEstimation}
E.~Castillo, A.~J. Conejo, R.~E. Pruneda, C.~Solares, and J.~M. Menendez,
  ``$m-k$ robust observability in state estimation,'' \emph{IEEE Transactions
  on Power Systems}, vol.~23, no.~2, pp. 296--305, May 2008.

\bibitem{SparsestSDIAOnlyFlowMeas}
K.~C. Sou, H.~Sandberg, and K.~H. Johansson, ``On the exact solution to a smart
  grid cyber-security analysis problem,'' \emph{IEEE Transactions on Smart
  Grid}, vol.~4, no.~2, pp. 856--865, June 2013.

\bibitem{MinSparcityDIAPMUTAC}
Y.~Zhao, A.~Goldsmith, and H.~V. Poor, ``Minimum sparsity of unobservable power
  network attacks,'' \emph{IEEE Transactions on Automatic Control}, vol.~62,
  no.~7, pp. 3354--3368, July 2017.

\bibitem{MinSparcityDIAPMUACC}
------, ``A polynomial-time method to find the sparsest unobservable attacks in
  power networks,'' in \emph{Proc. American Control Conference (ACC)}, July
  2016, pp. 276--282.

\bibitem{TopologicalObs1}
G.~R. Krumpholz, K.~A. Clements, and P.~W. Davis, ``Power system observability:
  A practical algorithm using network topology,'' \emph{IEEE Transactions on
  Power Apparatus and Systems}, vol. PAS-99, no.~4, pp. 1534--1542, July 1980.

\bibitem{ObservabulityAlgorithmGTMatroid}
V.~H. Quintana, A.~Simoes-Costa, and A.~Mandel, ``Power system topological
  observability using a direct graph-theoretic approach,'' \emph{IEEE
  Transactions on Power Apparatus and Systems}, vol. PAS-101, no.~3, pp.
  617--626, March 1982.

\bibitem{ObservabilityGTAlgorithm}
A.~Bargiela, M.~R. Irving, and M.~J.~H. Sterling, ``Observability determination
  in power system state estimation using a network flow technique,'' \emph{IEEE
  Transactions on Power Systems}, vol.~1, no.~2, pp. 108--112, May 1986.

\bibitem{GraphTheoryWest}
D.~B. West, \emph{\BIBforeignlanguage{English}{Introduction to Graph Theory}},
  2nd~ed.\hskip 1em plus 0.5em minus 0.4em\relax Upper Saddle River, N.J:
  Prentice Hall, 2001.

\bibitem{liu1stdatainjection}
Y.~Liu, P.~Ning, and M.~Reiter, ``False data injection attacks against state
  estimation in electric power grids,'' \emph{ACM Transactions on Information
  and System Security (TISSEC)}, vol.~14, no.~1, pp. 1--33, May 2011.

\bibitem{MATPOWER}
R.~Zimmerman, C.~Murillo-Sanchez, and R.~Thomas, ``Matpower: Steady-state
  operations, planning, and analysis tools for power systems research and
  education,'' \emph{IEEE Transactions on Power Systems}, vol.~26, no.~1, pp.
  12--19, Feb 2011.

\bibitem{HittingSetProbPaper}
K.~Chandrasekaran, R.~Karp, E.~Moreno-Centeno, and S.~Vempala, ``Algorithms for
  implicit hitting set problems,'' in \emph{Proceedings of the Twenty-second
  Annual ACM-SIAM Symposium on Discrete Algorithms}, ser. SODA '11, 2011, pp.
  614--629.

\bibitem{Karp1972}
R.~M. Karp, \emph{Reducibility among Combinatorial Problems}.\hskip 1em plus
  0.5em minus 0.4em\relax Boston, MA: Springer US, 1972, pp. 85--103.

\end{thebibliography}

\vfill

\end{document}